\documentclass[12pt,headings=normal]{article}
\pdfoutput=1
\usepackage[utf8]{inputenc}

\usepackage[normalem]{ulem}
\usepackage{amsmath,amssymb}
\usepackage{slashed}
\usepackage{xcolor}
\usepackage{setspace}

\usepackage{graphicx}
\usepackage{cite}
\usepackage{pdflscape}
\usepackage{multirow}
\usepackage[thinlines]{easytable}
\usepackage{url}
\usepackage{braket}
\usepackage{subfigure}
\usepackage{longtable}
\usepackage{booktabs}

\newcommand{\eref}[1]{Eq.~\eqref{eq:#1}}

\newcommand{\aref}[1]{Appendix~\ref{app:#1}}
\newcommand{\sref}[1]{Section~\ref{sec:#1}}
\newcommand{\cref}[1]{Chapter~\ref{ch:#1}}

\newcommand{\nnl}{\nonumber \\}

\newcommand{\beq}{\begin{equation}} 
\newcommand{\eeq}{\end{equation}} 
\newcommand{\ba}{\begin{array}}  
\newcommand{\ea}{\end{array}} 
\newcommand{\bea}{\begin{eqnarray}}  
\newcommand{\eea}{\end{eqnarray} }  
\newcommand{\be}{\begin{eqnarray}}  
\newcommand{\ee}{\end{eqnarray} }  
\newcommand{\bal}{\begin{align}}
\newcommand{\eal}{\end{align}}   
\newcommand{\bi}{\begin{itemize}}  
\newcommand{\ei}{\end{itemize}}  
\newcommand{\ben}{\begin{enumerate}}  
\newcommand{\een}{\end{enumerate}}  
\newcommand{\bc}{\begin{center}}
\newcommand{\ec}{\end{center}} 
\newcommand{\bt}{\begin{table}}
\newcommand{\et}{\end{table}}  
\newcommand{\btb}{\begin{tabular}}
\newcommand{\etb}{\end{tabular}}

\newcommand{\cO}{{\mathcal O}} 
 
\newcommand{\cL}{{\mathcal L}} 
 
\newcommand{\cM}{{\mathcal M}}

\newcommand{\mpl}{M_{\mathrm Pl}}

\def\mpl{\, M_{\rm Pl}}

\def\ra{\rangle}
\def\la{\langle}  
\def\lb{\lbrack}
\def\rb{\rbrack}

\newcommand{\eps}{\epsilon}

\newcommand{\1}{{\bf 1}}
\newcommand{\2}{{\bf 2}}

\newcommand{\kb}{{\bf k}}

\newcommand{\jb}{{\bf j}}

\begin{document}

\begin{titlepage}

\begin{center}
\begin{spacing}{1.6}
{\LARGE 
Soft Matters, or the Recursions with Massive Spinors
} \end{spacing} %
\vspace{1.4cm}

\renewcommand{\thefootnote}{\fnsymbol{footnote}}
{Adam~Falkowski${}^1$ and  Camila~S.~Machado${}^2$}
\renewcommand{\thefootnote}{\arabic{footnote}}
\setcounter{footnote}{0}

\vspace*{.8cm}
\centerline{\em ${}^1$Universit\'{e} Paris-Saclay, CNRS/IN2P3, IJCLab, 91405 Orsay,
France }\vspace{1.3mm}
\vspace{.2cm}

\centering
\em ${}^2$PRISMA$^{+}$  Cluster of Excellence  Mainz Institute for Theoretical Physics,  \\
 Johannes Gutenberg-Universit\"at Mainz, 55099 Mainz, Germany\vspace{1.3mm}
\vspace{.2cm}
\vspace*{.2cm}

\end{center}

\vspace*{10mm}
\begin{abstract}\noindent\normalsize

We discuss recursion relations for scattering amplitudes with massive particles of any spin. They are derived  via a two-parameter shift of momenta, combining a  BCFW-type spinor shift with the soft limit of a massless particle involved in the process. The technical innovation is that spinors corresponding to {\em massive} momenta are also shifted.  Our recursions lead to a reformulation of the soft theorems. The well-known Weinberg's soft factors are recovered and, in addition, the subleading factors appear reshaped such that they are directly applicable to massive amplitudes in the modern on-shell language. Moreover, we obtain new results in the context of non-minimal interactions of massive matter with photons and gravitons. These soft theorems are employed for practical calculations of Compton and higher-point scattering.  As a by-product, we introduce a convenient representation of the Compton scattering amplitude for any mass and spin. 

\end{abstract}

\end{titlepage}
\newpage

\renewcommand{\theequation}{\arabic{section}.\arabic{equation}}

\tableofcontents

\section{Introduction}
\label{sec:intro}
\setcounter{equation}{0}

Soft theorems describe general properties of scattering amplitudes in the infrared (IR) regime where the  (four-)momentum of one  massless particle approaches zero, $p^\mu \rightarrow \epsilon p^\mu $ with $\epsilon \to 0$. 
In this limit it is possible to write down a recursion relation connecting the amplitude to a lower-point one with the soft particle removed.    
For example, for soft photons with helicity $h=\pm 1$ 
and soft gravitons with $h= \pm 2$ the recursions take the form
\begin{align}
\label{e:soft}
\cM_{n+1}^{\pm 1} = \left(
\frac{S^{(0)}_{\pm 1}}{\epsilon^2} 
+\frac{S^{(1)}_{\pm 1}}{\epsilon}   \right)\cM_{n} 
\, ,\qquad \cM_{n+1}^{\pm 2} = \left( \frac{S^{(0)}_{\pm 2}}{\epsilon^3} +\frac{S^{(1)}_{\pm 2}}{\epsilon^2}   +\frac{S^{(2)}_{\pm 2}}{\epsilon}  \right)\cM_{n} . 
\end{align}
$S^{(i)}_{\pm h}$ are called the soft factors. 
They are {\em universal}, in the sense that they do not depend on what other ``hard" particles are involved in the scattering. 

Since the seminal works of Low~\cite{Low:1958sn} and Weinberg~\cite{Weinberg:1965nx}, which explored the leading term in the soft momentum expansion, the importance of the soft theorems has been revealed through applications that span very different fields. 
In particular, soft theorems turn out to be intricately related  to the consistency conditions of the S-matrix such as gauge invariance, locality and unitarity \cite{Broedel:2014fsa,Rodina:2018pcb}. 
Furthermore, it was shown that they can be understood as Ward identities of asymptotic symmetries~\cite{Strominger:2017zoo}.  
They have also been relevant to understanding the landscape of effective field theories (EFTs) (see e.g. \cite{Cheung:2015ota,Cheung:2016drk, Cheung:2018oki,Elvang:2018dco,Low:2019ynd}). 

A new life into the study of soft theorems was brought by the emergence of on-shell methods for calculating scattering processes. 
These are founded on the spinor techniques, which make manifest the little group transformation properties of the amplitudes.      
An important step in this development was Ref.~\cite{Cachazo:2014fwa} by Cachazo and Strominger. Using the BCFW recursion relations~\cite{Britto:2005fq} they were able to extend the graviton soft theorems up to sub-subleading order at tree level.\footnote{%
In this paper we restrict the discussion to tree level.
Although the leading (Weinberg's) soft factor is not modified at loop level, the story for the sub-leading orders is more subtle~\cite{Bern:1998sv,He:2014bga,Sahoo:2018lxl}.}
Subsequently, the work by Elvang et al.~\cite{Elvang:2016qvq} extended on-shell soft theorems to a broad class of theories with massless particles.
This is achieved thanks to a modified recursion relation that combines double BCFW and  soft shifts. 
This approach leads to an elegant proof that various S-matrix consistency conditions, such as charge conservation and the equivalence principle, follow directly from Poincar\'{e} invariance, locality, and unitarity, without ever mentioning gauge invariance. 
Moreover, Ref.~\cite{Elvang:2016qvq} systematized the classes of effective operators that modify the soft theorems at the subleading order for photons and the sub-subleading order for gravitons.

Despite these rapid developments, all of the results derived with on-shell techniques are valid only for massless theories.\footnote{For soft theorems with massive particles using the effective action formalism, see e.g. \cite{Bianchi:2015lnw,AtulBhatkar:2018kfi,Sen:2017nim,Laddha:2017ygw,Chakrabarti:2017ltl}.} 
Meanwhile, a convenient spinor formalism for dealing with {\em massive} particles was introduced in~\cite{Arkani-Hamed:2017jhn}. 
There are several motivations to rewrite the soft theorems such that they can be directly applied to theories using the on-shell language and massive spinors. 
Perhaps the most pressing is the growing importance of the on-shell program in the context of black hole computations \cite{Kosower:2018adc,Guevara:2018wpp,Laddha:2018rle,Arkani-Hamed:2019ymq}. 
Another is the progress in on-shell formulation of massive EFTs, 
such as e.g. massive gravity~\cite{Bonifacio:2018vzv,Bonifacio:2018aon,Bonifacio:2019mgk,Falkowski:2020mjq}, or the  Standard Model EFT \cite{Shadmi:2018xan,Aoude:2019tzn,Durieux:2019eor}. 
More generally, evolution of on-shell tools to deal with massive theories may reveal unexpected simplicity of their amplitudes, much  as it happened for massless theories. 

In this paper, we establish soft recursion relations that are valid for general theories with massive particles of any spin, extending the results of \cite{Elvang:2016qvq}.
This is achieved by introducing a new kind of holomorphic spinor shift, where also {\em massive} spinors are shifted in addition to massless ones. 
This allows us to derive a general recursion relation that remains valid even when the soft particle is the only massless particle in the scattering process. 
Our formula can be applied to describe emission of a soft particle of any helicity  $|h| \leq 2$. 
In this paper we work out the resulting soft theorems for photons and gravitons.
These have the same general structure as in the standard derivation, however they can be directly applied to amplitudes written in term of massive and massless spinors. 
In particular, the subleading soft factors become differential operators acting on the massive spinors and on their little group indices.
We explicitly connect the subleading soft theorems to multipole expansion of the 3-point interaction between massive matter and photons/gravitons. 
As a corollary, we obtain a compact and elegant proof that gravitational dipole interactions of fundamental particles of any spin are inconsistent with the assumption of Poincar\'{e} invariance, locality, and unitarity.

We also discuss a couple of concrete applications of the soft recursion relations in massive theories. 
One is for the simple 4-body process of Compton scattering of photons and gravitons.
For any spin of the matter particle, the recursion correctly captures the IR part of the amplitude, that is the one containing the physical poles in the Mandelstam variables. 
The result agrees with the one in Ref.~\cite{Arkani-Hamed:2017jhn} obtained by requiring that the residues of the physical poles factorize into 3-point amplitudes. 
As discussed in \cite{Arkani-Hamed:2017jhn,Chung:2018kqs}, for higher spins the IR part inevitably develops a spurious pole that is not associated with an exchange of a physical particle.
Therefore, to make it consistent with unitarity, the Compton amplitude has to be augmented by a UV term that cancels the spurious pole without affecting the physical ones.
In our derivation, this UV term arises as a boundary term in the recursion relation. 
As an aside, we  present a systematic algorithm to build the UV term in Compton scattering, alternative to the one in \cite{Chung:2018kqs}.  
We also make some baby steps toward higher-point amplitudes of any spin, by calculating specific 5-point amplitudes with 3 photons or 3 gravitons.  

Finally, we touch on the subject of exponential representation of soft amplitudes. 
Exponentiated soft factors play an important role in the matching between quantum and classical limits. 
In particular, they are instrumental in certain calculations of black hole observables~\cite{Guevara:2018wpp,Guevara:2019fsj}. 
They can also be used to write scattering amplitudes in the Mellin space, which can be connected with CFT correlators in the celestial sphere (see e.g. \cite{Cheung:2016iub,Pasterski:2017kqt,Donnay:2018neh}).
Our recursion relations lead to a simple proof of the exponentiation of Compton amplitudes~\cite{Guevara:2018wpp} for any mass and spin. 
We also remark that the photon and graviton soft factors can be written in terms of an exponential operator in the scattering of massive particles of generic spin, which previously was only done for the minimal-helicity violating (MHV) sector of massless gravity/Yang-Mills~\cite{He:2014bga}.

Our paper is organized as follows. The (massless and massive) spinor formalism is briefly reviewed in \sref{spinor}.
The new soft recursions, valid for particles of any mass and spin, are derived in \sref{soft}, 
from which the soft theorems in \sref{st} directly follow. 
Some applications for calculating concrete amplitudes are presented in \sref{app}.
\sref{exp} discusses the exponential representation of soft amplitudes and Compton scattering.  
\sref{conc} is reserved for conclusions and future directions.
\aref{shifts} lists the possible spinor shifts realizing our momentum shift, some of which are not employed in this paper but may be useful for other applications. 
\aref{uvCompton} describes in detail our construction of UV/boundary terms in Compton scattering.

\section{Spinor conventions}
\label{sec:spinor}
\setcounter{equation}{0}

We begin with a brief overview of our conventions (readers well-versed in the massless and massive spinor formalism are invited to skip this section).
We work in four dimensions with the mostly-minus metric $\eta_{\mu \nu} = {\rm diag}(1,-1,-1,-1)$. 
The Lorentz algebra can be decomposed into $SL(2,\mathbb{C}) \times SL(2,\mathbb{C})$. 
Holomorphic and antiholomorphic spinors $\psi_\alpha$ and $\psi_{\dot \alpha}$  transform under the respective $SL(2,\mathbb{C})$ factors with indices being raised and lowered by the antisymmetric epsilon tensor: 
\beq
\label{eq: spinordef}
\psi^\alpha = \eps^{\alpha \beta} \psi_\beta \ ,
\qquad 
\psi_\alpha = \eps_{\alpha \beta} \psi^\beta \quad, 
\eeq 
and idem for dotted indices.
Vector and spinor Lorentz indices can be traded with the help of the sigma matrices  $(\sigma^\mu)_{\alpha \dot \beta}  = (\mathbb{I}, \vec  \sigma )$ and 
$(\bar \sigma^\mu)^{\dot \alpha \beta}  = (\mathbb{I}, -\vec  \sigma )$, where $\vec \sigma$ are the  Pauli matrices. 
Given the momentum\footnote{%
In this paper ``momentum" always means four-momentum.} 
$p_\mu$, we can construct the $2 \times 2$ matrices 
\bea
p_\mu (\sigma^\mu)_{\alpha \dot \alpha} \equiv (p\sigma)_{\alpha \dot \alpha}, 
\qquad 
p_\mu (\bar \sigma^\mu)^{\dot \alpha \alpha} \equiv (p\bar \sigma)^{\dot \alpha \alpha}. 
\eea
These naturally act on the spinor indices: 
$(p\bar \sigma) \psi \equiv (p\bar \sigma)^{\dot \alpha \beta } \psi_\beta$, 
$(p\sigma) \tilde \psi \equiv (p\sigma)_{ \alpha \dot \beta } \tilde \psi^{\dot\beta}$,
$\tilde \psi (p \bar \sigma)  \equiv 
\tilde \psi_{\dot \beta} (p\bar \sigma)^{\dot \beta \alpha } $, 
$\psi (p\sigma)  \equiv \psi^\beta (p\sigma)_{ \beta \dot \alpha }$, 
where repeating spinor indices are implicitly summed over. 
We will often omit the spinor indices,
as with these rules the contractions are always unambiguous. 

{\bf Massless momentum} can be represented by a pair of (commuting) spinors $\lambda_\alpha$, $\tilde \lambda_{\dot \beta}$: 
\beq
\label{eq:SHF_lambdadef}
(p  \sigma)_{\alpha \dot \beta} =  \lambda_\alpha \tilde \lambda_{\dot \beta}, 
\qquad (p  \bar \sigma)^{\bar \alpha \beta}   =  \tilde \lambda^{\dot \alpha} \lambda^\beta\,,  
\eeq 
such that $p^2 = 0$ is automatic for any $\lambda$, $\tilde \lambda$ thanks to the identity $\lambda^\alpha \lambda_\alpha = 0 = \tilde \lambda_{\dot \alpha } \tilde \lambda^{\dot \alpha}$. 
It follows that the massless spinors satisfy the Weyl equations:  
$\lambda (p \sigma) =   (p \sigma)  \tilde \lambda = 0$, 
$(p \bar \sigma) \lambda = \tilde \lambda  (p \bar \sigma) = 0$. 
For real $p_\mu$ we have the additional constraint $\tilde \lambda = \bar \lambda$, but in general  momenta are allowed to be complex, in which case $\tilde \lambda$ and $\lambda$ are independent.  
The {\em little group} transformations (i.e. for a fixed $p_\mu$, the subset of Lorentz transformations leaving $p_\mu$ invariant) correspond  to $U(1)$ acting as $\lambda \to t^{-1} \lambda$, $\tilde \lambda \to t \tilde \lambda$. 
The Lorentz-invariant and little group covariant building blocks are  commonly represented by the bra-ket notation:
\beq
 \langle i j \rangle \equiv \lambda_i^\alpha \lambda_{j \, \alpha} =  \eps^{\beta \alpha } \lambda_{i \, \alpha} \lambda_{j\, \beta} =  (\lambda_i \lambda_j) , 
\qquad  
  [ i j ] \equiv \tilde \lambda_{i\, \dot \alpha} \tilde \lambda^{\dot \alpha}_j   = \eps^{\dot \alpha  \dot\beta} \tilde \lambda_{i \, \dot \alpha} \lambda_{j \, \dot \beta} 
  =  ( \tilde \lambda_i \tilde \lambda_j) \, , 
\eeq  
and the Lorentz contraction of momenta can be written as $2p_i p_j = \la ij \ra \lb ji\rb$.\footnote{%
We always follow the conventions of Ref.~\cite{Dreiner:2008tw}. 
For the holomorphic contraction $\langle \cdot \rangle$
this differs by a sign from some of the on-shell literature. } 
The bra-ket notation can be naturally extended to more complicated contractions, e.g. $
(\lambda_i p_k \sigma \tilde \lambda_j) \equiv \langle i  p_k j]$, 
$(\lambda_i p_k \sigma p_l \bar \sigma  \tilde \lambda_j)
\equiv \langle  i  p_k  p_l j]$, etc. 
In the following, we choose to write uncontracted spinors explicitly as in Eq.~(\ref{eq: spinordef}), and reserve the bra-ket notation for Lorentz contractions only; 
e.g. we write $\langle  i j \rangle \lambda_k \equiv 
(\lambda_i \lambda_j) \lambda_k $. 
The massless spinors can be related to polarization tensors in the standard Lagrangian formalism.  In particular, the polarization vectors of a massless spin-1 particle can be written in terms of the spinors as  
\beq
\label{eq:PT_epsilon_masslessspin1} 
\epsilon^-_\mu  =  { (\lambda \sigma_\mu \tilde  \zeta) \over \sqrt 2 \lb \lambda  \zeta \rb }, 
\qquad 
\epsilon_\mu^+    =  { (\zeta \sigma_\mu \tilde  \lambda) \over \sqrt 2 \la \lambda  \zeta \ra }, 
\eeq 
where $\tilde \zeta$ and $\zeta$ are arbitrary  reference spinors representing  the gauge freedom. 
Similarly, for a massless spin-2 particle, the polarization tensors are 
$ \epsilon^\pm_{\mu \nu} = \epsilon^\pm_\mu \epsilon^\pm_\nu$, 
where now the freedom of choosing the reference spinors represents general coordinate invariance.  

{\bf Massive momentum} satisfying the on-shell condition $p^2 = m^2$ can be represented by four spinors $\chi_{\alpha}^{\, 1}$, $ \chi_{\alpha}^{\, 2}$,  $\tilde \chi_{\dot \beta \, 1}$, $\tilde \chi_{\dot \beta \, 2}$.
They can be collected into  $\chi^J$ and $\tilde \chi_J$, where $J=1,2$ is identified with the $SU(2)$ little group index which, in complete analogy to spinor indices, can be raised and lowered by epsilon tensors. The spinors are related to the momentum by the formula 
\beq
\label{eq:SHF_chidef}
(p  \sigma)_{\alpha \dot \beta} =  \chi_{\alpha}^{\, J} \tilde \chi_{\dot \beta \, J}, 
\qquad (p  \bar \sigma)^{\alpha \dot \beta} =   \tilde \chi^{\dot \alpha}_{\, J} \chi^{\beta \, J}\,, 
\eeq 
where summation over the little group indices is implicit. 
This is a natural and convenient generalization of the spinor helicity formalism to massive particles~\cite{Arkani-Hamed:2017jhn} (for earlier works, see e.g.~\cite{Dittmaier:1998nn,Conde:2016vxs}). 
The massive spinors are normalized as 
\beq
\label{eq:SHF_chinorm}
(\chi^J \chi_K) = \delta^J_K  m ,   \qquad    
(\tilde \chi_J \tilde \chi^K) =   \delta_J^K  m, 
\eeq 
from which follow the Dirac equations 
\beq
\label{eq:SHF_chieom}
(p \sigma) \tilde \chi^J = m  \chi^J, 
\qquad 
(p  \bar \sigma )  \chi^J  =   m  \tilde \chi^J  , 
\qquad 
\chi^J (p  \sigma)  = -  m \tilde \chi^J, 
\qquad 
 \tilde \chi^J  (p\bar\sigma) = - m  \chi^J . 
\eeq 
These allow one to trade $\chi$ for $\tilde \chi$ and vice-versa.
An amplitude describing scattering of $n$ massive incoming particles with spin $S_i$, $i = 1 \dots n$,  for each $i$ will contain exactly $2 S_i$ spinors $\chi_i^{J_k}$ or $\tilde \chi_i^{J_k}$ with {\em uncontracted} little group indices. 
Massive {\em outgoing} particles can be represented by spinors $\chi_{i \, J_k}$ or  $\tilde \chi_{i \, J_k}$ with lowered little group indices, 
however in this paper we treat all external particles as incoming.  
The little group indices  corresponding to the same particle are always implicitly symmetrized over. 
To reduce clutter, in the following we will often omit the little group indices whenever it does not lead to ambiguities.  
Lorentz contractions of massive spinors will be shortened via the bra-ket notation:    
\beq
 \langle \mathbf{i j} \rangle \equiv  
 \chi_i^\alpha \chi_{j \, \alpha} = 
 (\chi_i \chi_j) , 
\qquad  
 [ \mathbf{i j}  ] \equiv 
 \tilde \chi_{i\, \dot \alpha} \tilde \chi^{\dot \alpha}_j  
  =  (\tilde \chi_i \tilde \chi_j) \, , 
\eeq  
where we adhere to the {\bf bold} notation introduced in Ref.~{\bf \cite{Arkani-Hamed:2017jhn}}.

\section{Soft Recursions}
\label{sec:soft}
\setcounter{equation}{0}

In this section we derive a formula connecting an $n+1$-point amplitude with at least one massless particle to $n$-point amplitudes. 
The formula picks up the leading terms in the limit where the momentum of a massless particle is continuously taken to zero. That massless particle is called {\em soft}, and the formula is referred to as the {\em soft recursion}.     
The derivation follows similar steps as the one in Ref.~\cite{Elvang:2016qvq}, however it is adapted to allow for massive particles in the scattering process.  

\subsection{Momentum shift}

Consider a set of $n+1$ particles with momenta $p_0 \dots p_n$ satisfying $\sum_{i=0}^n p_i = 0$ and the on-shell conditions $p_i^2 = m_i^2$.
The particle labeled as $0$ is always massless, $p_0^2=0$, and it will be our soft particle. 
We perform a two-parameter deformation of this kinematics.  
One parameter $z$ controls complex deformation of the momenta of the particle $0$ and of two other particles, below labeled as $j$ and $k$. The other parameter $\epsilon$ controls the soft limit of the particle 0. 
Specifically, the shift is defined as\footnote{%
The original reference~\cite{Elvang:2016qvq} defines the shift in a slightly different way: 
$\hat p_j = p_j - \left [  \epsilon  \alpha_j  - z \beta_j \right ]  q_j$, 
$\hat p_k = p_k + \left [ \epsilon   \alpha_k -  z \beta_k \right ]  q_k$, 
such that $\hat p_{j,k} = p_{j,k}$ for $\epsilon= z =0$. 
In this version the unshifted kinematics is constrained as $\sum_{i=1}^n p_i = 0$ (as opposed to $\sum_{i=0}^n p_i = 0$  in our case). For this reason we find it more convenient to work with  the version in \eref{ST_ss}. } 
\beq
\label{eq:ST_ss}
\hat p_0 = \epsilon p_0 - z q_0, \quad 
\hat p_j = p_j - 
{(\epsilon-1) (p_0 p_k)  - z (q_0 p_k) \over q_j p_k }  q_j , \quad
\hat p_k = p_k - 
{ (\epsilon-1)(p_0 p_j) -  z (q_0 p_j) \over q_k p_j}  q_k, 
\eeq
where $q_0$, $q_j$, $q_k$ are four-vectors satisfying 
\beq 
\label{eq:ST_ss_qcondition}
q_0^2 = q_j^2 = q_k^2 =  q_0 p_0 = q_j p_j = q_k p_k = 0, 
\qquad 
 q_j p_0 = q_k p_0  = q_j q_0  = q_k q_0 = q_j q_k  = 0. 
\eeq 
The first set of equations in \eref{ST_ss_qcondition} ensures that the shift does not change the particles' masses, $\hat p_i^2 = m_i^2$, 
while the role of the second is to ensure shifted momentum conservation,
$\sum_{i=0}^n \hat p_i = 0$, for any $z$ and $\epsilon$. 
One advantage of the two-parameter shift in \eref{ST_ss} compared to simpler BCFW-like one-parameter shifts is that it will directly lead to soft expansion of recursion relations in powers of $1/\epsilon$.
As a bonus, it will allow for more control over the recursions' boundary terms, as the latter will start at higher orders in $1/\epsilon$.

\subsection{Recursion} 

Using the momentum shift in \eref{ST_ss} one can relate, in a quite model-independent way,  $n+1$-point amplitudes in the soft limit to  $n$-point amplitudes. 
Consider an $n+1$ point amplitude $\cM_{n+1} \equiv \cM(1 \dots n 0)$. 
We denote $\hat \cM^{z,\epsilon}_{n+1}$ its shifted version, which is  $\cM_{n+1}$ evaluated for the ``hatted" kinematics in \eref{ST_ss}. 
Similarly, we denote $\cM_{n} \equiv \cM(1 \dots n)$, and  its shifted version by $\hat \cM_n^{z,\epsilon}$. 
Note that $\hat \cM^{0,1} = \cM$.
At tree level $\hat \cM^{z,\epsilon}$ is a meromorphic function of $z$, with all singularities given by a finite set of simple poles at $z = z_i$.
Applying the Cauchy formula, 
\beq
\label{eq:ST_cauchy}
\hat \cM^{0,\epsilon}_{n+1} = -\sum_i { {\rm Res}_{z \to z_i} \hat \cM^{z,\epsilon}_{n+1} \over z_i}  + B_\infty , 
\eeq 
where the boundary term is $B_\infty  = {1 \over 2 \pi i} \int_{C_\infty} {\hat \cM^{z,\epsilon}_{n+1} \over z}$, and $C_\infty$ is the circle at infinity.  If $\hat \cM^{z,\epsilon}_{n+1}$ goes to zero as $z \to \infty$ then  $B_\infty = 0$. 
The recursion relation is most powerful when $B_\infty = 0$, but this will not always be true in the cases of interest. 

We only focus on a subset of the poles related to the emission of the particle zero from an external leg the $n$-point amplitude  $\cM_n$.
In the limit $\epsilon \to 0$,  singular $1/\epsilon^n$ terms in $\hat \cM^{z,\epsilon}_{n+1}$ can only come from these residues. 
There is $n$ such poles $z_l$ related to the solutions of the equations $\hat P_{0l}^2|_{z = z_l}= m_l^2$, where $\hat P_{0l} \equiv \hat p_0 + \hat p_l$  and  it is understood that $ \hat p_l = p_l$ for $l \neq j,k$.
The poles are located at 
 \beq
\label{eq:ST_zl}   
z_l = \epsilon {p_0 p_l \over q_0 p_l} , 
\qquad l=1 \dots n,  
\eeq 
and one can rewrite 
$\hat P_{0l}^2 -  m_l^2 = -2 q_0 p_l (z - z_l)$. 
This shows that the poles at $\hat P_{0l}^2 \to  m_l^2$ are in one-to-one correspondence with the poles at $z \to z_l$. 
Unitarity requires that at every  pole the amplitude factorizes: \beq
\hat \cM^{z,\epsilon}(1 \dots n 0)|_{\hat P_{0l}^2 \to m_l^2}  = 
- {\hat \cM^{z_l,\epsilon}(1 \dots P_{0l} \dots  n) \hat \cM^{z_l,\epsilon}((-P_{0l}) l 0 )  \over 
\hat P_{0l}^2  -  m_l^2 } 
= { \hat \cM^{z_l,\epsilon}(1 \dots P_{0l} \dots  n) \hat \cM^{z_l,\epsilon}((-P_{0l}) l 0)  \over 
2 q_0 p_l (z - z_l) }  .  
\eeq  
Thus 
\beq
 {{\rm Res}_{z \to z_l} \hat \cM^{z,\epsilon}(0 \dots n)  \over z_l} 
 =  {1 \over \epsilon} { \hat \cM^{z_l,\epsilon}(1 \dots P_{0l} \dots  n) \hat \cM^{z_l,\epsilon}((-P_{0l}) l 0 )  \over 
2 p_0 p_l } . 
\eeq  
Plugging this back into the Cauchy formula in  \eref{ST_cauchy} on obtains the soft recursion: 
\beq
\label{eq:ST_recursion}
\hat \cM^{0,\epsilon}(1 \dots n 0) =  - {1 \over \epsilon } \sum_{l = 1}^n 
{ \hat \cM^{z_l,\epsilon}(1 \dots P_{0l} \dots  n) \hat \cM^{z_l,\epsilon}( (-P_{0l}) l  0 )  \over 2 p_0 p_l }  + \cO(\epsilon^0). 
\eeq
This formula will be the starting point for deriving a new on-shell incarnation of soft theorems which is valid in the presence of massive particles.  
We will often refer to the $n$-point amplitude above as the {\em hard factor}.  The boundary term in \eref{ST_cauchy} is absorbed in $\cO(\epsilon^0)$ above, as it cannot produce singular terms in the $\epsilon \to 0$ limit.  

\subsection{Spinor shift}

We turn to discussing spinor shifts that realize the momentum shift in \eref{ST_ss}. 
The discussion depends on whether the shifted particles $j$ and $k$ are massive or massless.
In the following we first work out the case when $m_j > 0$ and  $m_k > 0$, and later comment on other possible configurations. 
Note that $m_j$ and $m_k$ can be different.  
Recall also that particle $0$ is always massless in this discussion, however  other particles with $l \neq 0,j,k$ can be massive or massless, 
as one pleases.

One possible choice of the shift vectors $q$ in \eref{ST_ss} is  
\beq
\label{eq:ST_0mmbar_qsigma}
q_0 \sigma = y \tilde \lambda_0, \qquad 
q_j \sigma = p_j \sigma \tilde \lambda_0 \tilde \lambda_0, 
\qquad 
q_k \sigma = p_k \sigma \tilde \lambda_0 \tilde \lambda_0,   
\eeq
where $y$ is an arbitrary spinor satisfying $\la y 0 \ra\neq 0$. One can verify that these satisfy the conditions in \eref{ST_ss_qcondition}.
We can decompose the shifted momenta into spinors: $\hat p_0 \sigma = \lambda_0^z \tilde \lambda_0$, 
$\hat p_j \sigma =\chi_j^J \tilde \chi_{j \, J}^z$, 
$\hat p_k \sigma =\chi_k^J \tilde \chi_{k \, J}^z$. 
At the spinor level, the shift is realized as\footnote{%
In a different context, single parameter shifts of massive spinors were previously considered in Refs.~\cite{Aoude:2019tzn,Franken:2019wqr}.} 
\bea
\label{eq:ST_0mmbar_spinor}
{\rm \bf \{ 0 \bar m \bar m \} }: \qquad  
\lambda_{0}^z & = &  \eps  \lambda_0 - z  y, 
\nnl 
\tilde \chi_j^z & = & \tilde \chi_j  
+  {  (\epsilon -1 )  \la 0 p_k 0\rb   - z \la y  p_k 0 \rb  \over  \lb 0 p_j p_k 0 \rb }   \lb \jb  0\rb  \tilde \lambda_0 , 
 \nnl 
 \tilde \chi_k^z & = &
\tilde \chi_k   
 -  {  (\epsilon -1 )  \la 0  p_j  0 \rb   - z \la y  p_j 0 \rb  \over \lb 0  p_j p_k 0 \rb } \lb \kb 0 \rb \tilde \lambda_0 , 
\eea
while all other spinors remain unshifted. 
Since we shift the holomorphic spinor corresponding to the particle $0$ and the antiholomorphic spinors corresponding to the particles $j,k$, we label this shift as $\{ 0\bar m \bar m \}$. 
The parity-reversed version, labelled as $\{ \bar 0 m m \}$, is defined in \aref{shifts}. 
Note that our shift is non-trivial for $z = \epsilon = 0$; 
in this limit the spinors $\tilde \chi_{j,k}$ must shift to absorb the original momentum of the particle $0$. 
This ``zero-order" shift will play some role in the following. 
On the other hand, the shift is trivial for  $z =0$ and  $\epsilon = 1$, but that is away from the soft limit $\epsilon \to 0$ on which we are focused in this paper.

We end this section with a number of identities  that will be useful in the following (impatient readers are encouraged to skip to the next section). 
For the $\{ 0 \bar m \bar m \}$ shift,  the location of the poles in \eref{ST_zl} can be recast as  
\beq
z_l =  \epsilon { \langle 0 p_l 0]\over \la y p_l 0 \rb } \, ; 
\qquad 
m_l = 0 \  \Rightarrow \ z_l = \epsilon { \langle 0 l \rangle  \over \langle y l \rangle} . 
\eeq
At the poles, the shifted spinors can be expressed as \bea
\label{eq:ST_0mmbar_uf}
\lambda_0^{z_l }  & = &   \epsilon { \la y 0 \ra \over \la y p_l 0 \rb} p_l \sigma \tilde \lambda_0 , 
\nnl 
\tilde  \chi_j^{z_l} & = &  \tilde    \chi_j  
-  {2 p_0 p_k  
\over \lb 0 p_j p_k 0\rb } 
\lb \jb 0 \rb  \tilde    \lambda_0 
+ \epsilon  { \la y  0 \ra  \lb 0  p_k p_l 0\rb 
 \over \la y  p_l 0\ra \lb 0 p_j p_k 0\rb } 
 \lb \jb  0 \rb \tilde \lambda_0 , \nnl 
\tilde  \chi_k^{z_l} & = &  \tilde    \chi_k
+  {2 p_0 p_j  
\over \lb 0 p_j p_k 0\rb }
\lb \mathbf{k} 0\rb  \tilde    \lambda_0 
- \epsilon  { \la y  0 \ra  \lb 0  p_j p_l 0\rb 
 \over \la y  p_l 0\ra \lb 0 p_j p_k 0\rb } 
 \lb \mathbf{k} 0\rb \tilde \lambda_0 .
\eea 
These identities allow us to rewrite  $\hat P_{0 l}$ at the poles. 
For $l \neq j,k$, $\hat P_{0 l} = \hat p_0 + p_l$, 
thus 
\beq 
\hat P_{0l} \sigma |_{z_l} =  p_l \sigma 
+  \epsilon { \la y  0 \ra \over \la y p_l 0\rb}    p_l \sigma \tilde \lambda_0 \tilde \lambda_0, 
\qquad l \neq j,k. 
\eeq 
Since $\hat P_{0l}|_{z=z_l}$ is on shell by definition,  we can factorize it into spinors: 
\bea 
m_l > 0 & \,\Rightarrow \, & 
\hat P_{0l} \sigma |_{z_l} = \chi_{0l}^L \tilde \chi_{0l \, L}, \qquad \chi_{0l}  =  \chi_l, \quad 
\tilde \chi_{0l} =    \tilde \chi_l +   \epsilon { \la y  0 \ra   \over \la y p_l 0\rb} 
\lb \mathbf{l} 0\rb  \tilde \lambda_0,  \quad l\neq j,k, 
\nnl 
m_l = 0 & \,\Rightarrow \, & 
\hat P_{0l} \sigma |_{z_l} = \lambda_{0l} \tilde \lambda_{0l},
\qquad 
\lambda_{0l}  =  \lambda_l, \quad 
\tilde \lambda_{0l}  = \tilde  \lambda_l 
+ \epsilon { \la y  0 \ra \over \la y l\ra } \tilde \lambda_0 ,  \quad l\neq j,k . 
\eea 
Similarly at $z_j$ and $z_k$ we have the decomposition 
\bea
\hat P_{0j} \sigma|_{z_j} &= &  p_j \sigma 
- { 2 p_0 p_k \over  \lb 0 p_j p_k 0\rb} p_j \sigma \tilde \lambda_0  \tilde \lambda_0 
\ \Rightarrow \ \chi_{0j}  =  \chi_j, \quad 
\tilde \chi_{0j}  =    \tilde \chi_j 
  - {2 p_0 p_k
  \over  \lb 0 p_j p_k 0\rb} 
  \lb \mathbf{j} 0 \rb  \tilde \lambda_0, 
\nnl 
\hat P_{0k}  \sigma |_{z_k} &= &  p_k  \sigma  
 + {2 p_0 p_j \over  \lb 0 p_j p_k 0\rb} 
 p_k \sigma \tilde \lambda_0 \tilde \lambda_0, 
\ \Rightarrow \
 \chi_{0k}  =  \chi_k, \quad   \tilde \chi_{0k}  =    \tilde \chi_k 
+ {2 p_0 p_j \over  \lb 0 p_j p_k 0\rb} 
\lb \kb 0 \rb  \tilde \lambda_0. 
\qquad 
\eea 
In our story it will be important that these are independent of the soft parameter $\epsilon$. 
Similarly, the shifts $\tilde \chi_j^{z_k}$ and $\tilde \chi_k^{z_j}$ and are independent of $\epsilon$. 

The case where one of the $j,k$ particles is massive and the other is massless  is covered in \aref{shifts}.
We also comment on the limit where all involved particles are massless.  
This can be easily obtained from \eref{ST_0mmbar_spinor} by ``unbolding" the massive spinors, which corresponds to replacing $\chi_i^J \to \lambda_i$,  $\tilde \chi_i^J \to \tilde \lambda_i$ while dropping the $SU(2)$ little group index. 
Simplifying the resulting expression one obtains 
\bea
{\rm \bf \{ 0\bar 0 \bar 0 \} }:  \quad 
\lambda_{0}^{z}  & = &  \epsilon  \lambda_0 - z  y, 
\nnl 
\tilde \lambda_j^z  & = & \tilde \lambda_j  -  
{ (\epsilon - 1) \langle 0k \rangle -  z  \langle y k\rangle \over \langle j k \rangle } \tilde  \lambda_0,  
 \nnl 
 \tilde \lambda_k^z  & = & \tilde \lambda_k   +
{ (\epsilon - 1) \langle 0j \rangle -  z  \langle y j\rangle \over \langle j k \rangle } \tilde  \lambda_0, 
\eea
This reproduces the shift in Ref.~\cite{Elvang:2016qvq} up to the replacement $\epsilon \to \epsilon-1$ which is due to the different prescription for taking the soft limit.

\section{Soft Theorems}
\label{sec:st}
\setcounter{equation}{0}

In this section we apply the recursion formula in  \eref{ST_recursion} to derive the soft factors for amplitudes containing particles of any mass and spin. This generalizes the discussion in Ref.~\cite{Elvang:2016qvq}. 
To our knowledge, this is the first derivation using the on-shell methods with massive spinors.  
The soft factors that we obtain are applicable to more general physical situations than those found in the previous  literature. 
In the following we will deal with the soft particle of spin 2 (graviton) and spin 1 (photon), 
however \eref{ST_recursion} goes just as well with massless particles of spin 0 (Goldstone bosons), spin 1/2 (Goldstone fermions), or spin 3/2 (gravitinos).  
We restrict to single soft particle emission, leaving multiple soft limits for future publications. 

\subsection{Minimal gravity}

We begin with a gravity theory where the massless spin-2 graviton  is minimally coupled to matter and to itself.
The soft theorems in this case have been studied back and forth,  both in the standard Lagrangian and in the on-shell formalisms. 
Nevertheless, here we provide the soft theorems in a novel form that will be particularly convenient for practical  calculations with massive spinors.

In the on-shell formalism, the minimal coupling of the graviton  $h$ to a matter particle $X$ of mass $m>0$ and any spin $S$ corresponds to the on-shell 3-point amplitudes~\cite{Arkani-Hamed:2017jhn}
\bea 
\label{eq:GR_Mxxh_spinS_minimal}
\cM(\mathbf{1}_X \mathbf{2}_X 3_h^-) =  
- {1 \over \mpl } {\la 3 p_1 \zeta  \rb^2 \over \lb 3 \zeta \rb^2 } 
{ [\mathbf{21}]^{2S}   \over m^{2S}}, 
\qquad 
\cM(\mathbf{1}_X \mathbf{2}_X 3_h^+) =  
- {1 \over \mpl } {\la \zeta p_1 3 \rb^2 \over \la 3  \zeta \ra^2 } 
 {\langle \mathbf{21} \rangle^{2S}   \over m^{2S}}, 
\eea
where $\mpl = (8 \pi G)^{-1/2}  \approx 2.4 \times 10^{18}$~GeV is the Planck constant, and $\zeta$,  $\tilde \zeta$ are arbitrary reference spinors satisfying $\lb 3 \zeta \rb \neq 0$ and  $\la 3  \zeta \ra \neq 0$  (the amplitude is independent of their precise choice).
In the standard QFT language, the on-shell amplitude in \eref{GR_Mxxh_spinS_minimal} corresponds to graviton coupling to matter via the energy-momentum tensor. 
Meanwhile, the minimal self-coupling of the graviton in the  standard Einstein-Hilbert Lagrangian
corresponds in the on-shell language  to 
\beq
\label{eq:GR_Mhhh}
\cM(1_h^- 2_h^- 3_h^+) =  -  {1 \over \mpl}{\langle 12 \rangle^6 \over \langle 13 \rangle^2 \langle 23 \rangle^2 }, 
\qquad 
\cM(1_h^+ 2_h^+ 3_h^-) =   
-{1 \over \mpl }{[12]^6 \over [13]^2 [23]^2 }. 
\eeq 
We focus on emission of a soft graviton with plus helicity. 
The starting point is the recursion in \eref{ST_recursion}, with the amplitudes deformed by the $\{0 \bar m \bar m \}$ spinor shift in \eref{ST_0mmbar_spinor}.  
We first need an expression for the shifted 3-point amplitude at the $z_l$-poles:  
\beq
\label{eq:ST_hatM_minimal}
\hat \cM^{z_l,\epsilon} ( (-P_{0l})  l  0^+_h )   =  
- {1 \over \mpl} {\la y p_l 0 \rb^2 \over \epsilon^2 \la 0 y \ra^2 }      . 
\eeq
This expression is independent of the identity, mass, or spin of the $l$-th particle (except that for massive particles the identity operator in the little group indices is implicit), which is a realization of the equivalence principle.  
Plugging this back into \eref{ST_recursion} we obtain a raw version of the soft theorem:   
\beq 
\label{eq:ST_softGravity}
\hat \cM^{0,\epsilon}(1 \dots n 0^+_h) =  {1 \over  \epsilon^3 \mpl  } \sum_{l = 1}^n 
 { \la y p_l 0 \rb^2  \over (2 p_0 p_l) \la 0 y \ra^2  }  
  \hat \cM^{z_l,\epsilon}(1 \dots P_{0l} \dots  n)
+ \cO(\epsilon^0). 
\eeq 
To proceed we need to expand the $n$-point amplitude above in powers of $\epsilon$. 
Using the identities in \eref{ST_0mmbar_uf} we find 
\beq 
\label{eq:ST_epsilonExpansion}
\hat \cM^{z_l,\epsilon}(1 \dots P_{0l} \dots  n)   = 
\bigg \{  
1 +  \epsilon { \la 0 y \ra \over \la y p_l 0 \rb } \tilde  {\cal D }_l   
 +   {\epsilon^2  \over 2 } {\la 0 y \ra^2  \over \la y p_l 0 \rb^2   }  \tilde  {\cal D }_l ^2
\bigg \}  \hat \cM^{0,0}(1 \dots  n) +  \cO(\epsilon^3), 
\eeq 
where the differential operator $\tilde {\cal D}_l$ is 
\beq 
\label{eq:ST_calDl}
\tilde {\cal D }_l =  [0 \mathbf{l}] [0 \mathbf{\partial_l}]
+ { [0 p_k p_l 0] \over [0 p_j p_k 0] }  [0\mathbf{j}] [0 \mathbf{\partial_j}]
 -  { [0 p_j p_l 0] \over [0 p_j p_k 0] }   [0\mathbf{k}] [0 \mathbf{\partial_k}], 
 \quad   l \neq j,k, 
\eeq  
and $\tilde {\cal D }_l = 0$ for $l = j,k$. 
Here, $ [0 \mathbf{l}] [0 \mathbf{\partial_l}] \equiv 
\left (\tilde \lambda_0 \tilde \chi^L_l \right) \left (\tilde \lambda_0 { \partial \over \partial\tilde \chi_l^L} \right )$
for a massive particle $l$, where for once we explicitly displayed the $SU(2)$ little group index $L$, which is implicitly summed over.  
For a massless particle $l$ one should instead use 
$ [0 l] [0 \partial_l] \equiv 
\left (\tilde \lambda_0 \tilde \lambda_l \right) \left (\tilde \lambda_0 { \partial \over \partial\tilde \lambda_l} \right )$. Plugging the expanded $\hat \cM^{z_l,\epsilon}$  back into \eref{ST_softGravity} we obtain the soft theorem in the standard form:  
\beq 
\label{eq:ST_softTheoremGravity}
\hat \cM^{0,\epsilon}(1 \dots n 0_h^+ )  =   \bigg \{ 
{1 \over \epsilon^3 } S_{+2}^{(0)}  + {1 \over \epsilon^2 } S_{+2}^{(1)}   + {1 \over \epsilon} S_{+2}^{(0)}
\bigg \}   \hat \cM^{0,0}(1 \dots  n) 
+ \cO(\epsilon^0) .  
\eeq 
The leading soft factor is given by 
\beq
\label{eq:ST_leadingSFgravity}
S_{+2}^{(0)}  =   {1 \over \mpl}  \sum_{l=1}^n 
{ \la y p_l 0 \rb^2  \over  (2 p_0 p_l) \la 0 y \ra^2  } 
=  {1 \over \mpl} \epsilon_{\mu \nu}^{0+}   \sum_{l=1}^n 
{ p_l^\mu p_l^\nu \over p_0 p_l }.  
\eeq 
In the second we step we rewrote the soft factor in the more familiar form as it appears in Ref.~\cite{Weinberg:1965nx}, 
with 
the polarization tensor of the soft graviton defined below \eref{PT_epsilon_masslessspin1} and the gauge parameter $\zeta$ is identified with the shift spinor $y$. 
This leading soft factor is independent of $y$ as a consequence of the equivalence principle and momentum conservation~\cite{Weinberg:1964ew}: 
\beq 
{\partial \over \partial y } S_{+2}^{(0)} \sim \sum_{l=1}^n 
{    \la y p_l 0 \rb  \over \la y  0 \ra^3    }    y
= - {    \la y p_0  0\rb  \over \la y  0 \ra^3    }    y = 0. 
\eeq 
Conversely, the condition that $S_{+2}^{(0)}$ must be independent of the shift spinor $y$ can be employed to {\em prove} the equivalence principle without any reference to general coordinate invariance.\footnote{%
One might however argue that, in the on-shell approach to soft theorems, the role of the shift spinor $y$ is exactly the same as that of the gauge parameter in the standard approach. 
From this point of view, gauge invariance thrown out of the door, comes back in through the window. 
On the other hand, Ref.~\cite{Elvang:2016qvq} derives the same conclusion by demanding the absence of double poles in $z$, in which case the parallel with gauge invariance is avoided. }
Indeed, in our derivation, the soft theorems are a direct consequence of Poincar\'{e} invariance, locality, and unitarity. 
In a consistent theory they must lead to unambiguous soft factors for any choice of $y$. 
However, that would not be the case if the graviton coupling strength to  matter in \eref{GR_Mxxh_spinS_minimal}  were not universal and precisely correlated with the graviton self-coupling in \eref{GR_Mhhh}. 

The subleading soft factor in \eref{ST_softTheoremGravity} is a differential operator:
\beq
\label{eq:ST_subleadingSFgravity}
S_{+2}^{(1)}  =  {1 \over \mpl}   \sum_{l=1}^n 
{ \la y p_l 0 \rb  \over  (2 p_0 p_l) \la 0 y \ra  } \tilde {\cal D }_l .
\eeq 
It is  independent of the shift spinor $y$ as a consequence of  angular momentum conservation. 
Indeed,  
\beq
{\partial \over \partial y } S_{+2}^{(1)}  \sim \sum_{l=1}^n 
{ \la y  0 \ra p_l \sigma \tilde \lambda_0   
- \la y p_l 0 \rb  \lambda_0  \over  (2 p_0 p_l) \la y  0 \ra^2   } \tilde {\cal D }_l  
= - \sum_{l=1}^n 
{ \la 0 p_l   0\rb y    \over  (2 p_0 p_l) \la y  0 \ra^2   } \tilde {\cal D }_l 
= 
 - {  y  \over  \la y  0 \ra^2  } \sum_{l=1}^n  \tilde {\cal D }_l   . 
\eeq  
Now, $\sum_{l=1}^n  \tilde {\cal D }_l$ can be related to the angular momentum operator $J^{\mu \nu}$ introduced in \cite{Witten:2003nn} and  generalized to massive particles in \cite{Guevara:2018wpp}: 
\bea
\label{eq:QED_angularMomentum}
J^{\mu \nu}_l &=&  J_{\chi_l}^{\mu \nu} + J_{\tilde \chi_l}^{\mu \nu},   
\nnl 
J_{\chi_l}^{\mu \nu}   &\equiv & {i \over 4 } \left ( 
\chi_l  \sigma^\mu  \bar \sigma^\nu {\partial \over \partial \chi_l} 
- \chi_l \sigma^\nu  \bar \sigma^\mu {\partial \over \partial \chi_l}  \right ) , \qquad 
J_{\tilde \chi_l}^{\mu \nu}   \equiv {i \over 4 } \left ( 
\tilde  \chi_l   \bar \sigma^\mu  \sigma^\nu {\partial \over \partial \tilde  \chi_l} 
- \tilde  \chi_l  \bar \sigma^\nu  \sigma^\mu {\partial \over \partial \tilde  \chi_l}  \right ) . \quad 
\eea 
For a massless particle $l$ one should replace $\chi \to \lambda$. Trading  $\tilde {\cal D }_l$ for $J^{\mu \nu}$ we find 
\beq
{\partial \over \partial y }S_{+2}^{(1)}  \sim 
i  {(y \sigma^\mu \tilde \lambda_0) y  \over  \la y  0 \ra^3  }  p_0^\nu  \sum_{l =0}^n J^{\mu \nu}_l = 0 .  \eeq 
Finally, the sub-subleading soft factor in \eref{ST_softTheoremGravity} is a  double-differential operator:  
\beq
S_{+2}^{(2)} = {1 \over 4 \mpl } \sum_{l=1}^n 
{ 1 \over  p_0 p_l } \tilde {\cal D }_l^2   .  
\eeq 
This one is manifestly independent of the shift spinor $y$.  

Following the same steps, but starting instead with the $\{\bar 0 mm\}$ shift defined in \eref{ST_0barmm_spinor}, one can obtain the soft factors for emission of a minus helicity graviton:  
\beq
\hspace{-0.05cm}S_{-2}^{(0)}  =   {1 \over \mpl}   \sum_{l=1}^n 
 { \la 0 p_l y \rb^2    \over (2 p_0 p_l) \lb 0 y \rb^2  }, 
 \quad 
S_{-2}^{(1)}   =     {1 \over \mpl}   \sum_{l=1}^n 
 { \la 0 p_l y \rb    \over (2 p_0 p_l) \lb 0 y \rb  } {\cal D }_l  ,
 \quad 
S_{-2}^{(2)}  =    {1 \over 4 \mpl}  \sum_{l=1}^n 
 { 1   \over  p_0 p_l } {\cal D }_l^2 ,
\eeq 
where
\beq 
{\cal D }_l =  \langle 0 \mathbf{l}  \rangle   \langle 0 \mathbf{\partial_l} \rangle 
+ {  \langle 0 p_k p_l 0 \rangle   \over   \langle 0 p_j p_k 0 \rangle   }   \langle 0\mathbf{j} \rangle   \langle 0 \mathbf{\partial_j}\rangle  
 -  {  \langle 0 p_j p_l 0 \rangle   \over   \langle 0 p_j p_k 0 \rangle }   
 \langle 0\mathbf{k} \rangle  \langle 0 \mathbf{\partial_k} \rangle , 
~~   l \neq j,k~\text{and}~ {\cal D }_l =0, ~~ l=j,k.
\eeq 
Let us briefly comment on the relation of our results with the previous literature. 
It is a simple exercise to show that in the limit where all particles (also $j$ and $k$) are massless, the soft factors we derived reduce to those found in~\cite{Elvang:2016qvq}. 
In particular, the differential operator 
$\tilde {\cal D}_l$ reduces to 
\beq
\label{eq:ST_Dl_limit}
\tilde {\cal D}_l \to 
 [0 l] \left \{  [0 \partial_l]
 - {\langle l k \rangle \over \langle j k \rangle } 
 [0 \partial_j]
 +   {\langle l j \rangle \over \langle j k \rangle } 
 [0 \partial_k]  \right \} = 
 [0 l]  \nabla_{0l}, 
\eeq 
where $\nabla_{0l}$ is defined in Eq.~(19) of~\cite{Elvang:2016qvq}.   
On the other hand, our soft theorems do not exactly reduce to those in~\cite{Cachazo:2014fwa}, 
the obstruction being the last two terms in the curly bracket in  \eref{ST_Dl_limit}.
As discussed in~\cite{Elvang:2016qvq}, this difference is due to working with ``stripped" amplitudes $\cM$, rather than with $S$ matrix elements $T \sim \delta^4 (p_0 + \dots + p_N) \cM$ as in~\cite{Cachazo:2014fwa}. 
The price to pay for us is that the soft theorems look somewhat more complicated, and that they depend on the exact manner in which  the soft limit is taken.
The advantage is that for practical purposes one works with the stripped amplitudes $\cM$. 
The expressions we provide are ready to use out of the box with massive amplitudes written using the modern on-shell spinor formalism.

\subsection{Non-minimal gravity}

We move to consider theories where gravitational interactions of matter are non-minimal. 
Let us start by assuming that a massive particle $X$ with spin  $S > 0$ and mass $m>0$ has an anomalous gravitomagnetic dipole moment:  
\beq 
\label{eq:ST_GRdipole}
\cM(\mathbf{1}_X \mathbf{2}_X 3_h^+) \stackrel{?}{=}  
- {1 \over \mpl m^{2S} } {\la \zeta p_1 3 \rb^2 \over \la 3  \zeta \ra^2 } 
\left [   
 \langle \mathbf{21} \rangle^{2S}    
 +  {d_X \over m^2}    {\la \zeta p_1 3 \rb \over \la 3  \zeta \ra }  
  \langle 3 \mathbf{1} \rangle
    \langle 3 \mathbf{2} \rangle
 \langle \mathbf{21} \rangle^{2S-1}   
  \right ] .  
\eeq 
It was argued in~Ref.~\cite{Chung:2018kqs} that general coordinate invariance forbids  $d_X \neq 0$. 
But let us press on to see what happens.
It turns out that the dipole affects the subleading 
soft factor, which is modified as 
\beq
S_{+2}^{(1)}  = {1 \over \mpl}  \sum_{l=1}^n 
{ \la y p_l 0 \rb  \over  (2 p_0 p_l) \la 0 y \ra  } 
\left [    \delta^{\vec J}_{\vec L}   \tilde {\cal D }_l  
+  {d_l \over m_l^2}   \delta^{J_2 \dots J_{2 S_l} }_{L_2 \dots L_{2S_l}}   ( \tilde \chi_l^{J_1}  \tilde  \lambda_0)    (  \tilde \chi_{l \, L_1}  \tilde  \lambda_0)  
 \right ] . 
\eeq 
The first term in the bracket comes from the minimal coupling, and is the same as in \eref{ST_subleadingSFgravity}, except this time we explicitly display the identity operator acting on the little group indices of the $l$-th particle. 
The second factor is due to the dipole, where, as always, symmetrization of $J_1 \dots J_{2 S_l}$ and $L_1 \dots L_{2 S_l}$ is implicit. 
The problem is that, for $d_l \neq 0$, this soft factor is {\em not} independent of the shift spinor: 
\beq
{\partial \over \partial y } S_{+2}^{(1)}   \sim 
{  y  \over  \la y  0 \ra^2  } \delta^{J_2 \dots J_{2S_l} }_{L_2 \dots L_{2 S_l}}   \sum_{l=1}^n  {d_l \over m_l^2}  
  ( \tilde \chi_l^{J_1}  \tilde  \lambda_0)   (  \tilde \chi_{l \, L_1}  \tilde  \lambda_0) \neq 0.  
\eeq  
This provides a simple and general proof that, for a particle of any mass and spin, 
the presence of an anomalous gravitomagnetic dipole moment is in conflict with the assumptions of  Poincar\'{e} invariance, locality and unitarity.  

Given the dipole is forbidden, 
for $S \geq 1$ the leading deformation of the minimal coupling is the quadrupole: 
\bea 
\label{eq:GR_Mxxh_spinS_quadrupole}
\cM(\mathbf{1}_X \mathbf{2}_X 3_h^+) &= &  
- {1 \over \mpl m^{2S} } {\la \zeta p_1 3 \rb^2 \over \la 3  \zeta \ra^2 } 
\left [   
\langle \mathbf{21} \rangle^{2S}    
 +  {Q \over 2 m^4}    
 {\la \zeta p_1 3 \rb^2 \over \la 3  \zeta \ra^2 }  
  \langle 3 \mathbf{1} \rangle^2
    \langle 3 \mathbf{2} \rangle^2 
 \langle \mathbf{21} \rangle^{2S-2}   
+ \dots   \right ] ,
 \nnl 
\cM(\mathbf{1}_X \mathbf{2}_X 3_h^-) &= &  
- {1 \over \mpl  m^{2S} } 
{\la 3 p_1 \zeta  \rb^2 \over \lb 3 \zeta \rb^2 } 
\left [ [\mathbf{21}]^{2S}    
+ {Q \over 2 m^4}   {\la 3 p_1 \zeta  \rb^2 \over \lb 3 \zeta \rb^2 } 
 [ 3 \mathbf{1}]^2
   [ 3 \mathbf{2}]^2 
[ \mathbf{21}]^{2S-2}     
 + \dots \right ],
\eea
where the dots stand for eventual higher multipoles, 
which do not affect the soft theorems.\footnote{
The $n$-th multipole  enters the recursion in \eref{ST_recursion} with the 
${(\lambda_0^{z_l} \chi_1)^n (\lambda_0^{z_l} \chi_2)^n  \over (\lambda_0^{z_l} \zeta)^n }$ factor, which scales as $\epsilon^n$, given 
$\lambda_0^{z_l} \sim \epsilon$. } 
The quadrupole does arise from local Lagrangians, roughly speaking,  when gravity interacts with matter via the Riemann tensor;  
for example, for $S=1$ it corresponds to the coupling 
$\cL \sim Q X_{\mu \nu} X_{\alpha \beta} R_{\mu \nu \alpha \beta}$. 

We find that the quadrupole enters into the sub-subleading soft factor: \bea 
\label{eq:subsubleadingSFgravity_quadrupole}
S_{+2}^{(2)}  &= &  {1 \over 4 \mpl} \sum_{l=1}^n 
{ 1 \over  p_0 p_l  } \left [  
\tilde {\cal D }_l^2  
 \delta^{\vec J}_{\vec L}
+  {Q_l  \over m_l^2}     
 (\tilde \chi_l^{J_1} \tilde \lambda_0)    (\tilde \chi_l^{J_2} \tilde \lambda_0)  (\tilde \chi_{l \, L_1} \tilde \lambda_0)  
   (\tilde \chi_{l \, L_2} \tilde \lambda_0)  
   \delta^{J_3 \dots J_{2S_l} }_{L_3 \dots L_{2S_l}}    
 \right ] , 
\nnl  
S_-^{(2)}   &= &   {1 \over 4 \mpl}   \sum_{l=1}^n 
 { 1   \over  p_0 p_l} \left [   
 {\cal D }_l^2   \delta^{\vec J}_{\vec L}
+   { Q_l \over m_l^2}   
  (\chi_l^{J_1}  \lambda_0)    (\chi_l^{J_2}  \lambda_0)  ( \chi_{l \, L_1}  \lambda_0)  (\chi_{l \, L_2}  \lambda_0) 
  \delta^{J_3 \dots J_{2S_l} }_{L_3 \dots L_{2S_l}}   \right ] ,   \quad 
\eea
where, as always, symmetrization over the little group indices is understood. 
The double-differential operator due to the minimal coupling 
is now supplemented by an algebraic operator originating from the quadrupole. 
Note that the latter acts non-trivially on the little group indices: 
the $l$-th term in the sum acts on the upper indices 
$\vec L \equiv L_1 \dots L_{2S_l}$ of the $l$-th particle in $\hat \cM^{0,0}(1 \dots l^{\vec L} \dots  n)$.

\subsection{Photons}

Photons can be tackled by exactly the same methods as gravitons, except that the calculations are simpler.
Below we summarize the soft theorems for photon emission.  
Electromagnetic interactions of a particle $X$ with mass $m$ and spin $S$ are described by the following 3-point amplitudes~\cite{Arkani-Hamed:2017jhn,Chung:2018kqs}: 
\bea
\label{eq:QED_dipole}
\cM(\mathbf{1}_X \mathbf{2}_{\bar X} 3_\gamma^+) & = &
  {\sqrt 2   e \over m^{2S} }    \left \{    
 q {\la \zeta p_1 3 \rb \over \la 3  \zeta \ra }    \langle \mathbf{21} \rangle^{2S} 
+  {S a \over m^2} {\la \zeta p_1 3 \rb^2 \over \la 3  \zeta \ra^2 }  
       \langle 3 \mathbf{1} \rangle
    \langle 3 \mathbf{2} \rangle
 \langle \mathbf{21} \rangle^{2S-1}  
 + \dots  \right \} , 
     \nnl 
\cM(\mathbf{1}_X \mathbf{2}_{\bar X} 3_\gamma^-) & = &
  {\sqrt 2  e \over m^{2S} }  \left \{ 
 q {\la 3 p_1 \zeta  \rb \over \lb 3 \zeta \rb } [\mathbf{21}]^{2S}
   + {S \bar  a  \over   m^2} 
   {\la 3 p_1 \zeta  \rb^2 \over \lb 3 \zeta \rb^2 } 
 [ 3 \mathbf{1}]
   [ 3 \mathbf{2}] 
 [\mathbf{21}]^{2S-1} + \dots  \right \} , 
\eea  
where $e$ is the electromagnetic coupling constant, 
$q$ is the electric charge, 
$(a + \bar a)/2 = (g - 2)/2$ is the anomalous magnetic moment in the standard normalization,
$a \neq \bar a$ gives the electric dipole moment,   
and the dots stand for eventual higher multipoles which are irrelevant for the soft theorems.   
The first term in each bracket describes the minimal coupling, 
while the second describes non-minimal dipole interactions for $S \geq 1/2$. 
In the Lagrangian parlance, the former arises when photon interacts via 
$D_\mu = \partial_\mu - i e q A_\mu$ in the kinetic term, 
while the higher multipoles correspond to photon interacting via $F_{\mu \nu}$. 
For example,  for real $a$, the $S = 1/2$ Lagrangian would be 
$\cL \supset  e q  \bar \psi \gamma_\mu \psi   A_\mu
+ {e a \over 4 m} \bar \psi \sigma_{\mu \nu} \psi F^{\mu \nu}$, 
while for $S=1$ it would be 
$\cL \supset - i q e (V_{\mu \nu} \bar V_\nu - \bar V_{\mu \nu} V_\nu ) A_\mu
+ i e (q + 2 a) V_\mu \bar V_\nu F_{\mu \nu}$.  

Plugging the plus helicity amplitude in \eref{QED_dipole} into the recursion in \eref{ST_recursion}  one finds the raw version of the soft theorem: 
\bea 
\label{eq:ST_photonRaw}
\hat \cM^{0,\epsilon}(1 \dots n 0^+)  & = &    - { e \over \sqrt 2 } \sum_{l = 1}^n {1 \over p_0 p_l} \left [ 
{q_l \la y p_l 0 \rb  \over \epsilon^2  \la 0 y \ra }   \delta^{\vec J}_{\vec L} 
 +   {S_l a_l \over m_l \epsilon} 
 (\lambda_0 \chi_l^{J_1}) (\lambda_0 \chi_{l \, {L_1}})
   \delta^{J_2 \dots J_{2S_l} }_{L_2 \dots L_{2S_l}}  
\right ] 
  \hat \cM^{z_l,\epsilon}(1 \dots P_{0l}^{\vec L} \dots  n)
  \nnl & &  +\, \cO(\epsilon^0). 
\eea  
The first term in the bracket originates from the minimal coupling, while the second describes a correction from the dipole. 
The latter has a non-trivial action on the little group indices of  $\hat \cM$. 
Using the expansion of $\hat \cM$ in powers of $\epsilon$, cf. \eref{ST_epsilonExpansion}, one derives the soft theorem:  
\beq
\label{eq:ST_softTheoremPhotons}
\hat \cM^{0,\epsilon}(1 \dots n 0^\pm) = -   \bigg \{ 
{1 \over \epsilon^2 } S_{\pm 1}^{(0)}   
+ {1 \over \epsilon}S_{\pm 1}^{(1)}  
\bigg \}  \hat \cM^{0,0}(1 \dots  n)  
+ \cO(\epsilon^0) .
\eeq 
The leading soft factor for emission of a plus helicity photon
is 
\beq 
S_{+1}^{(0)}   = {e \over \sqrt 2}  \sum_{l = 1}^n  { q_l  \la y p_l 0 \rb  \over  \la 0 y \ra (p_0 p_l) }  
  = e \epsilon^{0+}_\mu \sum_{l = 1}^n { q_l p_l^\mu \over p_0 p_l},  \eeq 
where in the second step we introduced the polarization vector of the soft photon defined in \eref{PT_epsilon_masslessspin1},  
so as to make contact with the Weinberg's formula.   
The independence of $S^{(0)}$ on the shift spinor $y$ follows from charge conservation~\cite{Weinberg:1964ew}: 
\beq
{\partial  \over  \partial y}  S_{+1}^{(0)} 
\sim  {  y  \over  \la y  0 \ra^2}  
\sum_{l = 1}^n q_l   = 0;
\eeq 
or the other way around: 
given that the soft recursion must be valid for any $y$ proves charge conservation in any Poincar\'{e} invariant, local, and unitary theory.

The subleading soft factor is given by 
\beq 
\label{eq:subleadingSF_dipole}
S_{+1}^{(1)}  = 
 {e \over \sqrt 2 } \sum_{l = 1}^n {1 \over p_0 p_l} \left [ q_l \tilde  {\cal D}_l   \delta^{\vec J}_{\vec L} 
 +   {S_l a_l \over  m_l} 
 (\tilde \lambda_0 \tilde \chi_l^{J_1}) (\tilde \lambda_0 \tilde \chi_{l \, {L_1}})
   \delta^{J_2 \dots J_{2S_l} }_{L_2 \dots L_{2S_l}}  
\right ],  
\eeq 
where $\tilde  {\cal D }_l $ is defined in \eref{ST_calDl}. 
This one is manifestly independent of $y$, which  in particular confirms what everybody knows that arbitrary electromagnetic dipoles are allowed. 
Here, we display the action on little group indices of a massive $l$-th particle, which is non-trivial when dipoles are involved.
If on the other hand the $l$-th particle is massless, we should unbold the $l$-th term above, that is trade $\chi_l^L \to \lambda_l$,  dropping the $SU(2)$ little group indices.  
Finally, the soft factors for emission of a minus helicity photon are
\bea 
 S_{-1}^{(0)}  & = &    {e \over \sqrt 2 }   \sum_{l = 1}^n  { q_l   \langle 0 p_l y]    
  \over [0y ] ( p_0 p_l) }  
 = e \epsilon^{0-}_\mu \sum_{l = 1}^n { q_l  p_l^\mu \over p_0 p_l},   \nnl 
 S_{-1}^{(1)}  & = &  {e \over \sqrt 2 }  \sum_{l=1}^n {1 \over  p_0 p_l }\left [q_l  {\cal D}_l \delta^{\vec J}_{\vec L}  
  +   {S_l \bar a_l \over  m_l } 
 (\lambda_0 \chi_l^{J_1}) ( \lambda_0 \chi_{l \, {L_1}})
   \delta^{J_2 \dots J_{2S_l} }_{L_2 \dots L_{2S_l}} \right ]  .
 \eea

\section{Applications}
\label{sec:app}
\setcounter{equation}{0}

The goal of this section is to demonstrate that the recursion relations and soft theorems we obtained previously allow one to  efficiently calculate amplitudes with massive particles of arbitrary spin. 
Below we will reproduce some known results, as well as derive new ones not encountered in the literature so far.  

\subsection{Compton scattering}

We start with the Compton scattering.   
We are interested in the amplitude $\cM(\mathbf{1}_X \mathbf{2}_{\bar X} 3_\gamma^\pm 0_\gamma^+)$, where $0_\gamma$ is a soft photon\footnote{%
In the Mandelstam variables the soft photon limit $p_0 \to 0$  corresponds to 
$t \to m^2$, $u \to m^2$, $s \to 0$. 
For $m>0$ this of course cannot be realized with real kinematics, but the limit does make sense in complex kinematics. } 
of plus helicity, and $X$ is a particle of mass $m$, unit charge, and arbitrary spin $S$. 
We will use the $\{0 \bar m \bar m \}$ shift in \eref{ST_0mmbar_spinor} with $j=1$ and $k=2$.
We first assume minimal coupling, 
that is the absence of the dipole and higher multipoles in \eref{QED_dipole} (the dipole case will be discussed shortly). 
The soft theorem in \eref{ST_softTheoremPhotons} reduces to
\beq
\label{eq:softTheoremCompton}
\hat\cM^{0,\epsilon}(\mathbf{1}_X \mathbf{2}_{\bar X} 3_\gamma^\pm 0_\gamma^+)=
- {1 \over \epsilon^2} S_{+1}^{(0)}  
\hat \cM^{0,0}(\mathbf{1}_X \mathbf{2}_{\bar X} 3_\gamma^\pm)  
+ B_\infty.  
\eeq
In this case the subleading soft factor $S_{+1}^{(1)}$ is void for the simple reason that, for Compton scattering, we only have two factorization poles $z_{1,2}$, and  $\tilde {\cal D}_l$ in \eref{ST_calDl} is zero for $l=j,k$. 
Somewhat pedantically, we replaced $\cO(\epsilon^0)$ in \eref{ST_softTheoremPhotons} by the boundary term $B_\infty$, which is also  $\cO(\epsilon^0)$. 
The point is that, for 4-body processes, photon emission from external legs accounts for all factorization poles. 
Therefore, the recursion relation in \eref{ST_recursion} should reproduce the full Compton amplitude (not just the leading terms for $\epsilon \to 0$) up to the boundary term introduced in \eref{ST_cauchy}.
The leading soft factor for Compton scattering is easily calculated as 
\beq
S_{+1}^{(0)}   = \sqrt 2 e {\langle 3 p_1 0]  [03] 
\over (2p_1 p_0) (2 p_2 p_0)} . 
\eeq 
For minimal coupling, the hard 3-point amplitude 
on the right-hand side of \eref{softTheoremCompton} evaluates to 
\bea
\label{eq:APP_ComptonM00}
\hat \cM^{0,0}(\mathbf{1}_X \mathbf{2}_{\bar X} 3_\gamma^-) & = & 
 \sqrt 2 e {\langle 3 p_1 0 ] \over [30]}
 \left [
 - { \langle \mathbf{1} 3 \rangle [\mathbf{2} 0] +\langle \mathbf{2} 3 \rangle [\mathbf{1} 0]  \over  
 \langle 3 p_1 0] } \right ]^{2S}, 
\nnl 
\hat \cM^{0,0}(\mathbf{1}_X \mathbf{2}_{\bar X} 3_\gamma^+) & = & 
 \sqrt 2 e {[3 0] \over \langle 3 p_1 0] }
 { \langle \mathbf{21}\rangle ^{2S} \over m^{2S-2} }. 
\eea 
It is important to stress that the shift is not void  in $\hat \cM^{0,0}$.   
That is because the shift we defined in
\eref{ST_0mmbar_spinor} is non-trivial for $z = \epsilon = 0$ (it is trivial if $\epsilon=1$ instead). 
For this reason $\hat \cM^{0,0}(\mathbf{1}_X \mathbf{2}_{\bar X} 3_\gamma^\pm)$ above is not simply a copy of 
$\cM(\mathbf{1}_X \mathbf{2}_{\bar X} 3_\gamma^\pm)$ in \eref{softTheoremCompton}.
This zero-order shift is absolutely crucial to recover the correct result. 
In particular, the non-trivial spinor structure  in $\hat \cM^{0,0}(\mathbf{1}_X \mathbf{2}_{\bar X} 3_\gamma^-)$   appears as a consequence of the identity 
$(\tilde \chi_1^z \tilde \chi_2^z)|_{z=\epsilon=0} =  
m  { \langle \mathbf{1} 3 \rangle [\mathbf{2} 0] +\langle \mathbf{2} 3 \rangle [\mathbf{1} 0]  \over  
 \langle 3 p_1 0] } $. 
All in all we reconstruct 
\bea
\label{eq:APP_compton}
\hat\cM^{0,\epsilon}(\mathbf{1}_X \mathbf{2}_{\bar X} 3_\gamma^- 0_\gamma^+) 
& = & 
{2 e^2  \langle 3 p_1 0]^2 \over \epsilon^2 (2p_1 p_0) (2 p_2 p_0)} 
 \left [
 - { \langle \mathbf{1} 3 \rangle [\mathbf{2} 0] +\langle \mathbf{2} 3 \rangle [\mathbf{1} 0]  \over  
 \langle 3 p_1 0] } \right ]^{2S} + B_\infty , 
\nnl
\hat\cM^{0,\epsilon}(\mathbf{1}_X \mathbf{2}_{\bar X} 3_\gamma^+ 0_\gamma^+)
& = & 
{2 e^2 [3 0]^2  \over \epsilon^2 (2p_1 p_0) (2 p_2 p_0)}
 {\langle \mathbf{21}\rangle^{2S} \over m^{2S-2} } + B_\infty' . \eea 
On general grounds, this correctly describes leading terms of the Compton amplitude in the limit when the momentum of $0_\gamma^+$ is soft. 
The full amplitude valid for {\em any} momentum of $0_\gamma^+$ is recovered by setting $\epsilon=1$. 
In this limit, the same-helicity amplitude above is a fully consistent expression for any $S$, also when  $B_\infty' = 0$.\footnote{%
For $S>1$, the same-helicity Compton amplitude diverges as $(p_1 + p_2)^2 \to \infty$, in which case it is consistent in the sense of an effective theory amplitude below some cutoff scale.}
A non-zero $B_\infty'$  corresponds to a contact term (that is to say, without any singularities in the kinematic variables);  
for example we  can have $B_\infty'  =  {[\mathbf{21}]^{2S} \over m^{2S} }{[3 0]^2  \over \Lambda^2}$.   
On the other hand, the opposite-helicity amplitude is healthy as it stands only for $S \leq 1$. 
As discussed in \cite{Arkani-Hamed:2017jhn}, 
for $S>1$ the first term in \eref{APP_compton} develops an unphysical pole at 
$ \langle 3 p_1 0] \to 0$. 
That means the boundary term cannot vanish for $S>1$;  it must be non-zero to cancel the unphysical pole, and at the same time it cannot have any singularities as $p_0 p_1 \to 0$ or $p_0 p_2 \to 0$.    
Therefore it must be of the form  
$B_\infty = \sum_{k =1}^{2S-2} 
{C_k \over \langle 3 p_1 0]^k}$, 
where $C_k$ are pure contact terms.  

One possible form of the boundary term is 
\bea
\label{eq:QED_comptonS_UVeven}
B_\infty &  = &  
-  { 2 e^2  \over  (2 m^2)^{2S} (2 p_1 p_0) (2 p_2 p_0)}  
 \sum_{k= 3}^{2S}  \binom{2S}{k}  { 
 {\cal Y} \big ({\cal Y}^{k-1} - \{  {\cal Y} || {\cal Z} \} {\cal Z}^{k-2} \big ) (- {\cal X})^{2S - k} 
 \over 
\langle 3 p_1  0]^{k-2} }, 
\nnl 
{\cal Y} & \equiv & (2p_1 p_0) \langle \mathbf{1} 3 \rangle [\mathbf{2} 0] + (2 p_2 p_0) \langle \mathbf{2} 3 \rangle [\mathbf{1} 0], 
\qquad 
{\cal Z}  \equiv  (2p_1 p_0) \langle \mathbf{1} 3 \rangle [\mathbf{2} 0] - (2 p_2 p_0) \langle \mathbf{2} 3 \rangle [\mathbf{1} 0], 
\nnl 
{\cal X }    & \equiv  &   
m \langle \mathbf{12}  \rangle + m [\mathbf{12}]
-\langle \mathbf{1} p_0 \mathbf{2}] -\langle \mathbf{2} p_3 \mathbf{1}] ,  
\eea 
where $\{  {\cal Y} || {\cal Z} \}  =  {\cal Y} $ for even $k$, 
and $\{  {\cal Y} || {\cal Z} \}  =  {\cal Z} $ for odd $k$. 
Any other legal boundary term will differ from the one above only by contact terms.  
The absence of physical poles in \eref{QED_comptonS_UVeven} is not manifest, 
but it immediately follows from 
${\cal Y}^{2n} - {\cal Z}^{2n}= ({\cal Y}^2 - {\cal Z}^2)({\cal Y}^{2n-2} + \dots + {\cal Z}^{2n-2})$,
given that $({\cal Y}^2 - {\cal Z}^2) = 16 (p_1 p_0) (p_2 p_0)
\langle \mathbf{1} 3 \rangle [\mathbf{2} 0]  
\langle \mathbf{2} 3 \rangle [\mathbf{1} 0]$. 
A rationale for this boundary term is postponed to \aref{uvCompton}. For the present discussion we notice that $B_\infty$ is a polynomial  in $z$ after applying our $\{0 \bar m \bar m \}$ shift,
which is due to the fact that the  shift enters only via the $p_1 p_0$ and $p_2 p_0$ factors in 
${\cal Y}$ and ${\cal Z}$.
More precisely, 
the boundary term is of the form 
$B_\infty = \sum_{n=1}^{2S-2} {P_n(z) \over \langle 3 p_1  0]^n }$, 
where $P_n(z)$ is the  $n$-th order polynomial in $z$. 
Furthermore, $P_n(z) \sim \epsilon^n$ in the soft limit. 
Thus the boundary term does not show up in the soft theorems,  
but still it contributes to the recursion in \eref{ST_recursion} via the integral over the circle at infinity. 
Overall $B_\infty \sim \cO(\epsilon^1)$, 
better than  $\cO(\epsilon^0)$ deduced on general grounds.  
The boundary term we propose is arguably more compact than the one quoted in \cite{Chung:2018kqs}.  
It also makes manifest that the UV behavior of Compton scattering for $S>1$ can be $\cO(E^{4S}/m^{4S})$. 
In fact, one can further soften it down to  $\cO(E^{4S-2}/m^{4S-2})$ by subtracting from  it a judiciously chosen contact term. 
See \aref{uvCompton} for more details. 

We turn to another  example where the subleading soft factor is no longer moot. 
We consider Compton scattering for arbitrary $S \geq 1/2$ when the dipole $a$ in \eref{QED_dipole} is switched on. 
In order to keep this example simple,  we restrict to the case where $q = 0$, however a more general formula can be easily worked out by the same methods. 
The hard amplitude with a minus helicity photon is given by 
\beq
\hat \cM^{0,0}(1_X 2_{\bar X} 3_\gamma^-)
= (-1)^{2S+1} {\sqrt 2 e S \bar a  \over m}  \langle \mathbf{1} 3 \rangle  \langle \mathbf{2} 3 \rangle 
 \left [
 { \langle \mathbf{1} 3 \rangle [\mathbf{2} 0] +\langle \mathbf{2} 3 \rangle [\mathbf{1} 0]  \over  
 \langle 3 p_1 0] } \right ]^{2S-1}.  
\eeq  
The subleading soft factor for a plus helicity photon takes the form 
\beq
S_{+1}^{(1)}  =  { \sqrt 2 S e a  \over  m } \left [ 
{ [0 \mathbf{1}]
 (\tilde \lambda_0 \tilde \chi_{1 \, {L_1}})\over 2 p_0 p_1}
-
{  [0 \mathbf{2}]
 (\tilde \lambda_0 \tilde \chi_{2 \, {K_1}})\over 2 p_0 p_2}
\right ] .
\eeq 
Only those little group indices that act non-trivially on the hard factor are displayed.
With a little help of spinor algebra one can show that this action amounts to 
\bea 
 [0 \mathbf{1}] (\tilde \lambda_0 \tilde \chi_{1 \, {L_1}} ) 
\left \{ \langle \mathbf{1} 3 \rangle  \langle \mathbf{2} 3 \rangle 
 \bigg  (
 \langle \mathbf{1} 3 \rangle [\mathbf{2} 0] +\langle \mathbf{2} 3 \rangle [\mathbf{1} 0]    \bigg  )^{2S-1}
 \right \} & = &  
 -  \langle 3 p_1 0]  \langle \mathbf{2} 3 \rangle [\mathbf{1} 0]   
\bigg (  \langle \mathbf{1} 3 \rangle [\mathbf{2} 0] +\langle \mathbf{2} 3 \rangle [\mathbf{1} 0]    \bigg  )^{2S-2} 
\nnl & \times & 
\left [ \langle \mathbf{1} 3 \rangle [\mathbf{2} 0] + {1 \over 2 S} \langle \mathbf{2} 3 \rangle [\mathbf{1} 0]    \right ] ,
\nnl 
\,  [0 \mathbf{2}]
 (\tilde \lambda_0 \tilde \chi_{2 \, {K_1}}) 
 \left \{ \langle \mathbf{1} 3 \rangle  \langle \mathbf{2} 3 \rangle 
 \bigg (  \langle \mathbf{1} 3 \rangle [\mathbf{2} 0] +\langle \mathbf{2} 3 \rangle [\mathbf{1} 0]    \bigg )^{2S-1} 
  \right \}  
 & = &  
   \langle 3 p_1 0]  \langle \mathbf{1} 3 \rangle [\mathbf{2} 0]   
\bigg (   \langle \mathbf{1} 3 \rangle [\mathbf{2} 0] +\langle \mathbf{2} 3 \rangle [\mathbf{1} 0]    \bigg  )^{2S-2} 
\nnl & \times & 
\left [  {1 \over 2 S} \langle \mathbf{1} 3 \rangle [\mathbf{2} 0] + \langle \mathbf{2} 3 \rangle [\mathbf{1} 0]    \right ]. 
\eea  
The $1/2S$  factors appear as a consequence of symmetrization of the little group indices. 
Putting it all  together:
 \bea 
 \label{eq:APP_comptonDipole}
 & \displaystyle
\hat\cM^{0,\epsilon}(\mathbf{1}_X \mathbf{2}_{\bar X} 3_\gamma^- 0_\gamma^+) 
=   - {1 \over \epsilon} S_{+1}^{(1)}  \hat \cM^{0,0}(1_X 2_{\bar X} 3_\gamma^-)  +  B_\infty 
\nnl  & \displaystyle
=   |a|^2 {(-1)^{2S} S e^2  \over \epsilon m^2  (2p_1 p_0) (2 p_2 p_0)} 
 \left [
 { \langle \mathbf{1} 3 \rangle [\mathbf{2} 0] +\langle \mathbf{2} 3 \rangle [\mathbf{1} 0]  \over  
 \langle 3 p_1 0]  } \right ]^{2S-2} 
&  \nnl & \displaystyle \times   
 \bigg \{  (2 p_1 p_0)  \langle \mathbf{1} 3 \rangle [\mathbf{2} 0]
\big(   \langle \mathbf{1} 3 \rangle [\mathbf{2} 0] +2 S \langle \mathbf{2} 3 \rangle [\mathbf{1} 0]  \big  )  
+ 
(2 p_2 p_0)  \langle \mathbf{2} 3 \rangle [\mathbf{1} 0]
\big  (  2 S  \langle \mathbf{1} 3 \rangle [\mathbf{2} 0] +\langle \mathbf{2} 3 \rangle [\mathbf{1} 0]  \big ) 
 \bigg  \}   +  B_\infty  . \qquad  \eea 
This  describes the leading effect of the dipole in the limit where the momentum $p_0$ is soft.
Away from this limit, the expression makes sense for $B_\infty = 0$ as long as $S \leq 1$, 
whereas for $S \geq 3/2$ one needs to adjust $B_\infty$ so as to cancel the unphysical pole at $\langle 3 p_1 0]\to 0$. 
A possible boundary term can be constructed using the techniques discussed in \aref{uvCompton}, 
and is displayed in full glory in \eref{UV_dipoleBoundary}.

\subsection{Gravitational Compton scattering}

Calculation of gravitational Compton scattering amplitudes 
$\cM(\mathbf{1}_X \mathbf{2}_X 3_h 0_h)$ follows the steps laid out in the previous subsection, with only minor modifications.  
We again work with the $\{0 \bar m \bar m \}$ shift of the $012$ legs. 
Consider matter particles minimally coupled to gravity as in \eref{GR_Mxxh_spinS_minimal}. 
The leading soft factor $S_{+2}^{(0)}$  in \eref{ST_leadingSFgravity} is easily calculated as 
\beq
S_{+2}^{(0)} = - 
{\langle 3 p_1 0]^2  [03]^2  
\over \mpl (2p_1 p_0) (2 p_2 p_0) (2 p_3 p_0)} . 
\eeq 
The absence of subleading soft factors for minimal coupling is just a tad less trivial to demonstrate than in the photon case.
Starting from the expression in \eref{ST_subleadingSFgravity} one notes that for our shift $\tilde {\cal D}_1 = \tilde {\cal D}_2 = 0$, while the remaining one reduces to 
\beq
\tilde {\cal D}_3 = 
[0 \1] \left (\tilde \lambda_0 { \partial  \over \partial \chi_1 }\right ) + [0 \2] \left (\tilde \lambda_0 { \partial  \over \partial \chi_2 }\right ) + [0 3] \left (\tilde \lambda_0 { \partial  \over \partial \lambda_3 }\right ).
\eeq 
Thus, $\tilde {\cal D}_3$ is proportional to the total angular momentum operator, 
and it annihilates 3-point amplitudes. 
As in the photon case, the IR behavior of gravitational Compton amplitudes is fully determined by  the leading soft factor, as long as the graviton is minimally coupled to matter. Given \eref{GR_Mxxh_spinS_minimal} and the zero-order shift in \eref{ST_0mmbar_spinor}, 
the relevant hard factors are evaluated as  
 \bea
\label{eq:APP_comptonGravityM00}
\hat \cM^{0,0}(\mathbf{1}_X \mathbf{2}_X 3_h^-) & = & 
- {1 \over \mpl} {\langle 3 p_1 0 ]^2 \over [30]^2}
 \left [
 - { \langle \mathbf{1} 3 \rangle [\mathbf{2} 0] +\langle \mathbf{2} 3 \rangle [\mathbf{1} 0]  \over  
 \langle 3 p_1 0] } \right ]^{2S}, 
\nnl 
\hat \cM^{0,0}(\mathbf{1}_X \mathbf{2}_X 3_h^+) & = & 
- {1 \over \mpl} {[3 0]^2 \over 
\langle 3 p_1  0]^2 }
 { \langle \mathbf{21}\rangle ^{2S} \over m^{2S-4} }. 
\eea 
Thus, the gravitational Compton scattering amplitudes are reconstructed as 
\begin{align}
\label{eq:APP_comptonGravity}
&\hat \cM^{0,\epsilon}(\mathbf{1}_X \mathbf{2}_X 3_h^- 0_h^+)  =  
{S_{+2}^{(0)} \over \epsilon^3}\hat \cM^{0,0}(\mathbf{1}_X \mathbf{2}_X 3_h^-) 
= 
{(-1)^{2S} \langle 3 p_1 0]^4 \over \epsilon^3 \mpl^2 (2p_1 p_0) (2 p_2 p_0) (2 p_3 p_0)} 
 \left [ { \langle \mathbf{1} 3 \rangle [\mathbf{2} 0] +\langle \mathbf{2} 3 \rangle [\mathbf{1} 0]  \over  
 \langle 3 p_1 0] } \right ]^{2S}\!\!\!\!\!\! + B_\infty ,
\nnl 
&\hat \cM^{0,\epsilon}(\mathbf{1}_X \mathbf{2}_X 3_h^+0_h^+)  =  
{S_{+2}^{(0)} \over \epsilon^3}\hat \cM^{0,0}(\mathbf{1}_X \mathbf{2}_X 3_h^+) = 
{[3 0]^4  \over \epsilon^3 \mpl^2 (2p_1 p_0) (2 p_2 p_0)(2 p_3 p_0)}
{\langle \mathbf{21}\rangle^{2S} \over m^{2S-4} } 
+ B_\infty' .
\end{align}
The IR part agrees with the expression derived in \cite{Arkani-Hamed:2017jhn} by matching the residues at the kinematic poles. 
The amplitude away from the soft limit is obtained using 
$\cM = \hat \cM^{0,1}$.
Much as for photons, it is consistent to set  $B_\infty' = 0$ for any spin, however for $S>2$ the boundary term $B_\infty$ has to be non-zero so as to cancel the unphysical  $\langle 3 p_1 0]$ pole, see \aref{uvComptonGR} for an explicit expression.  

Let us also work out a simple example beyond the leading soft factor.
Consider a massive particle with spin $S \geq 1$ and non-minimal quadrupole interactions with the graviton, corresponding to $Q \neq 0$  in \eref{GR_Mxxh_spinS_quadrupole}. 
In this case, the plus-helicity sub-subleading soft factor in \eref{subsubleadingSFgravity_quadrupole} for gravitational Compton scattering reduces to 
\beq 
S_{+2}^{(2)} =  {Q \over 4 \mpl m^2}\bigg \{ 
{  [\mathbf{1} 0] ^2  (\tilde \chi_{l \, L_1} \tilde \lambda_0) (\tilde \chi_{l \, L_2} \tilde \lambda_0)   \over  p_0 p_1 }
+ {  [\mathbf{2} 0] ^2   (\tilde \chi_{2 \, K_1} \tilde \lambda_0) (\tilde \chi_{2 \, K_2} \tilde \lambda_0)   \over  p_0 p_2 }
\bigg \} ,
\eeq 
where we displayed the little group indices acting non-trivially on the hard factor.  
The latter for plus helicity photons is generalized as  
\beq
\hat \cM^{0,0}(\mathbf{1}_X \mathbf{2}_X 3_h^+) =  
- {1 \over \mpl m^{2S-4} }  \left \{ 
{ [3 0]^2 \over  \langle 3 p_1  0]^2 }  \langle \mathbf{21}\rangle ^{2S}
+ 
{Q \over 2} {[3 0]^4 \over  \langle 3 p_1  0]^4 }  
 \langle \mathbf{1}3\rangle^2   \langle \mathbf{2}3\rangle^2  \langle \mathbf{21}\rangle ^{2S-2}  
 \right  \}  . 
\eeq 
Applying this to derive, for example, the same-helicity Compton amplitude we find   
\bea 
\hat \cM^{0,\epsilon}(\mathbf{1}_X \mathbf{2}_X 3_h^+0_h^+) & = & 
\left [ {S_{+2}^{(0)} \over \epsilon^3} 
+ {S_{+2}^{(2)} \over \epsilon}
\right ] \hat \cM^{0,0}(\mathbf{1}_X \mathbf{2}_X 3_h^+) = 
{[3 0]^4  \over \epsilon^3 \mpl^2 (2p_1 p_0) (2 p_2 p_0)(2 p_3 p_0)}
{\langle \mathbf{21}\rangle^{2S} \over m^{2S-4} } 
\nnl &+ &
 {Q \over 2} {[3 0]^6  \over \epsilon^3 \mpl^2 (2p_1 p_0) (2 p_2 p_0)(2 p_3 p_0)}
 { \langle \mathbf{1}3\rangle^2   \langle \mathbf{2}3\rangle^2   \over 
 \langle 3 p_1  0]^2  } 
{\langle \mathbf{21}\rangle^{2S-2} \over m^{2S-4} } 
\nnl &- & 
{Q \over 2 }  {[3 0]^2  \over  \epsilon  \mpl^2  \langle 3 p_1 0]^2}
\left [ {  [\1 0] ^2 \langle \2 p_1  0 ]^2  \over 2 p_0 p_1 }
+ {  [\20] ^2 \langle \1 p_2 0 ]^2  \over  2 p_0 p_2 }
\right ]  {\langle \mathbf{21}\rangle^{2S-2} \over m^{2S-2} } 
+ \cO(Q^2)
+ B_\infty' .
\nnl 
\eea 
Unlike in the minimal case, in the presence of the quadrupole 
 the unphysical $\langle 3 p_1 0]$ pole appears also in the same-helicity amplitude, 
 and it is present already for $S=1$. 
If one uses this formula away from the soft limit, 
one needs to include an appropriate boundary term $B_\infty'$ to cancel this pole for any $S \geq 1$

\subsection{Higher-point}

To conclude this section, we discuss a couple of examples of 5-point amplitude calculations using the soft theorems of \sref{st}. 
The goal is to provide an illustration how these formulas should be employed in the case of higher-point amplitudes.  
The examples involve a matter particle $X$ with  mass $m$, arbitrary spin $S$, and the minimal coupling to photons and gravitons. 

Our first example is the amplitude describing the production of three photons in the $X\bar X$ annihilation:  
$\cM(\mathbf{1}_X \mathbf{2}_{\bar X} 3_\gamma 4_\gamma 0_\gamma)$. 
We use the $\{0 \bar m \bar m \}$ ($\{\bar 0 m  m \}$) shift on the $012$ legs to calculate emission of a soft photon $0_\gamma$ with plus (minus) helcity. 
The leading soft factors for emission of $0_\gamma$ are given by
\beq
S_{+1}^{(0)}   = { \sqrt 2 e [0 p_1 p_2  0] 
\over (2p_1 p_0) (2 p_2 p_0)}, 
\qquad 
S_{-1}^{(0)}   = { \sqrt 2 e \langle 0 p_1 p_2  0 \rangle  
\over (2p_1 p_0) (2 p_2 p_0)}. 
\eeq 
These expressions are in fact true for 
$\cM(\mathbf{1}_X \mathbf{2}_{\bar X} 3_\gamma \dots n_\gamma 0_\gamma)$ with any $n-2 \geq 1$ number of hard photons.
The subleading factors are moot for the minimal coupling. 
That is  because the only charged particles are at the positions $1$ and $2$,
however ${\cal D}_{1,2} = 0$ for the shift we have chosen.  
Given \eref{APP_compton}, 
the hard factor in the soft recursion is given by 
\beq 
\hat\cM^{0,0}(\mathbf{1}_X \mathbf{2}_{\bar X} 3_\gamma^+ 4_\gamma^+)
 =  
{2 e^2 [3 4]^2  \over (2 \tilde p_1 p_3) (2 \tilde p_1 p_4)}
 {\langle \mathbf{21}\rangle^{2S} \over m^{2S-2} }  , 
\eeq 
where 
\beq
2  \tilde p_1 p_l \equiv 2  \tilde  p_1 p_l|_{z = 0, \epsilon=0} =
2 p_1  p_l  + 2 p_2 p_0 {[0 p_1 p_l  0] \over [0 p_1 p_2  0] }.  \eeq 
Putting it all together, 
\bea
\hat\cM^{0,\epsilon}(\mathbf{1}_X \mathbf{2}_{\bar X} 3_\gamma^+ 4_\gamma^+ 0_\gamma^+)   & = & {2\sqrt 2 e^3  [3 4]^2  [0 p_1 p_2  0] 
\over \epsilon^2 (2p_1 p_0) (2 p_2 p_0)(2 \tilde p_1 p_3) (2 \tilde p_1 p_4) }  {\langle \mathbf{21}\rangle^{2S} \over m^{2S-2} } + \cO(\epsilon^0), 
\nnl
\hat\cM^{0,\epsilon}(\mathbf{1}_X \mathbf{2}_{\bar X} 3_\gamma^+ 4_\gamma^+ 0_\gamma^-)   & = & 
{ 2\sqrt 2 e^3  [3 4]^2  \langle 0 p_1 p_2  0 \rangle  
\over \epsilon^2 (2p_1 p_0) (2 p_2 p_0) (2 \tilde p_1 p_3) (2 \tilde p_1 p_4)}  {\langle \mathbf{21}\rangle^{2S} \over m^{2S-2} }+ \cO(\epsilon^0) . 
\eea 

The similar approach can be used when photons are replaced with gravitons: $\cM(\mathbf{1}_X \mathbf{2}_X 3_h 4_h 0_h)$. 
Below we sketch the derivation, however the final formulas are too lengthy to quote here. 
The leading soft factor for emission of a plus helicity graviton is 
\beq
S_{+2}^{(0)} = {1 \over \mpl \langle 0 3 \rangle^2} \bigg \{
{\langle 3 p_1 0]^2 \over 2 p_0 p_1}
+ {\langle 3 p_2 0]^2 \over 2 p_0 p_2}
+ {\langle 3 4 \rangle^2 [40]^2 \over 2 p_0 p_4}
\bigg \}. 
\eeq 
Here we set $y = \lambda_3$ in \eref{ST_leadingSFgravity}, 
since the expression is true for an arbitrary spinor $y$. 
A major difference compared to the photon calculation before is that now the subleading soft factors are non-trivial. 
Indeed, picking again  $y = \lambda_3$ in \eref{ST_subleadingSFgravity}, 
the subleading soft factor can be written as 
\beq
S_{+2}^{(1)} = {\langle 3 4 \rangle [40] \over \mpl (2 p_0 p_4) \langle 0 3 \rangle} \tilde {\cal D}_4, 
\qquad 
 \tilde {\cal D}_4 = 
  [0 4] [0 \partial_4]
+ {  \langle 4 p_2 0] [40] \over  [0 p_1 p_2  0] }  [0\mathbf{1}] [0 \mathbf{\partial_1}]
 -  {  \langle 4  p_1 0] [40] 
 \over  [0 p_1 p_2  0] }   [0\mathbf{2}] [0 \mathbf{\partial_2}].  
\eeq 
The soft expansion of the 5-point amplitude $\cM^{0,\epsilon}(\mathbf{1}_X \mathbf{2}_X 3_h^+ 4_h^+ 0_h^+)$ is calculated  using \eref{ST_softTheoremGravity} with the soft factors above acting on  $\cM^{0,0}(\mathbf{1}_X \mathbf{2}_X 3_h^+ 4_h^+)$,  which in turn can be obtained from \eref{APP_comptonGravity}.
One can explicitly verify that 
$ \tilde {\cal D}_4 \cM^{0,0}(\mathbf{1}_X \mathbf{2}_X 3_h^+ 4_h^+) \neq 0$, 
thus the subleading soft factor is not moot in this case.

\section{Exponential representation}
\label{sec:exp}
\setcounter{equation}{0}

It has been known for a long time that the multipole expansion of classical spinning bodies, such as a Kerr black hole, can be resumed in a simple exponential form \cite{Hansen:1974zz}. 
However, the corresponding description in terms of on-shell amplitudes was obtained just recently in~\cite{Guevara:2018wpp}. 
That reference noticed that the on-shell 3-point amplitude describing minimal graviton interactions with matter of any spin can be recast in a compact exponential form, with the angular momentum operator in the exponent.  
Subsequently, they demonstrated that also the 4-point gravitational Compton scattering amplitude can be recast in an exponential form. 
This formal rephrasing turns out to be  useful to compute the scattering angle of two spinning black holes~\cite{Guevara:2018wpp}. 
As we are going to show, our soft recursion leads to a simple proof of the exponentiation of the Compton amplitudes for arbitrary spin of the matter particle. 
In the following, we demonstrate it for the photon amplitudes, however the derivation for gravity is totally analogous. 
We also discuss the exponential representation of the soft theorems for any (integer) helicity of the soft particle.

\subsection{Photon exponentiation}

Let us first rewrite the photon's minimal coupling as  
\beq
\label{eq:QED_Mffa_massive_minimal} 
\cM(\mathbf{1}_X \mathbf{2}_{\bar X} 3_\gamma^-)  = 
{2 e q (p_1\epsilon_3^-) \over m^{2S} }  
\lb \2\1 \rb^{2S}, \qquad   
 \cM(\mathbf{1}_X \mathbf{2}_{\bar X} 3_\gamma^+)  =  
{2 e q (p_1 \epsilon_3^+)  \over m^{2S} }  
\la \2\1 \ra^{2S}    . 
\eeq 
This is obtained from the leading term in Eq.~(\ref{eq:QED_dipole}) by replacing the $\zeta$ terms with the polarization vectors defined in \eref{PT_epsilon_masslessspin1}.\footnote{While polarization vectors themselves depends on the gauge (or the reference spinor $\zeta$), the 3-point function is gauge-independent because $p_1 p_3 = 0$ on the 3-body kinematics.} 
\eref{QED_Mffa_massive_minimal} can be equivalently recast as 
\bea
\label{eq:QED_Mxxa_exponential} 
\cM(\mathbf{1}_X \mathbf{2}_{\bar X} 3_\gamma^-) & = &
 { 2 e q (p_1 \epsilon_3^-)  \over m^{2S} }  
  \exp \left ( - i { \epsilon_{3}^{- \, \mu} p_3^\nu J_1^{\mu\nu}  \over (p_1 \epsilon_3^-)} \right )
 \la \2\1\ra^{2S},
 \nnl  
\cM(\mathbf{1}_X \mathbf{2}_{\bar X} 3_\gamma^+) & = &
 {2 e q (p_1 \epsilon_3^+)  \over m^{2S} }
  \exp \left (-  i { \epsilon_{3}^{+ \, \mu} p_3^\nu J_1^{\mu\nu}  \over (p_1\epsilon_3^+)} \right )  
   \lb \2\1\rb^{2S} ,  
\eea  
where $J^{\mu \nu}_i$ are the angular momentum operators introduced in \eref{QED_angularMomentum}. 
Note that we could equivalently replace $1 \to 2$ in the exponent. 
The proof relies on the  identities
$\la \mathbf{12} \ra -  \lb \mathbf{12} \rb =   
{\la 3 1 \zeta \rb  \over m^2 \lb 3 \zeta \rb}  
[ 3 \mathbf{1} ][ 3 \mathbf{2} ] 
= {\la \zeta  1 3 \rb  \over m^2 [3 \zeta]}  
\langle 3 \mathbf{1} \rangle \langle 3 \mathbf{2} \rangle $ and 
\beq
\epsilon_{3}^{- \, \mu} p_3^\nu J_{\chi_l}^{\mu\nu}  = 
{i \over \sqrt 2} 
\la 3 \mathbf{l} \ra \left (\lambda_3  {\partial \over \partial  \chi_l} \right ),
\quad \epsilon_{3}^{+ \, \mu} p_3^\nu J_{\tilde \chi_l}^{\mu\nu}  = {i \over \sqrt 2} 
\lb 3 \mathbf{l} \rb \left (\tilde \lambda_{3 }  {\partial \over \partial  \tilde \chi_{l}} \right ), \quad
\epsilon_{3}^{- \, \mu} p_3^\nu J_{\tilde \chi_l}^{\mu\nu}  =
\epsilon_{3}^{+ \, \mu} p_3^\nu J_{\chi_l}^{\mu\nu}  = 0, 
\eeq
from which it follows 
$\chi_2 \left ( 1 - i { \epsilon_{3}^{-  \, \mu} p_3^\nu J_1^{\mu\nu}  \over (p_1 \epsilon_3^- )}\right ) \chi_1  =  \lb \2\1\rb$,
$\tilde \chi_2 \left ( 1 - i { \epsilon_{3}^{+ \, \mu} p_3^\nu J_1^{\mu\nu}  \over (p_1 \epsilon_3^+ )} \right )  \tilde  \chi_1    = \la \2\1 \ra$. 
 
We  move to the calculation of the opposite helicity Compton amplitude using the soft recursion in \eref{ST_recursion} with the  $\{0 \bar m \bar m\}$ shift on the 012 legs. 
For the soft 3-point amplitude we use the exponentiated form in \eref{QED_Mxxa_exponential}, while for the hard factor we use the minimal form in \eref{QED_Mffa_massive_minimal}. 
Then the recursion reduces to 
\bea
\label{eq:ST_exp_1}
\hspace{-1.2cm}\hat \cM^{0,\epsilon}(\mathbf{1}_X \mathbf{2}_{\bar X} 3_\gamma^- 0_\gamma^+)  &=  &  
{(-1)^{2S} 4 q^2 e^2 \over \epsilon m^{2S} } \bigg \{  
{(\hat p_2 \epsilon_3^-) (\hat p_1 \hat \epsilon_0^+) \over u - m^2} 
 \exp \left ( - i { \epsilon_{3}^{- \, \mu} p_3^\nu J_2^{\mu\nu}  \over (\hat p_2 \epsilon_3^-)} \right )   
 \la \2  \1 \ra^{2S}  |_{z=z_1}
\nnl & + &{ (\hat p_1 \epsilon_3^-) (\hat p_2 \hat \epsilon_0^+) \over t - m^2} 
   \exp \left (-  i { \epsilon_{3}^{- \, \mu} p_3^\nu J_1^{\mu\nu}  \over (\hat p_1 \epsilon_3^-)} \right )  
  \la \2  \1 \ra^{2S}  |_{z=z_2} \bigg \} + B_\infty,   
\eea 
where we already contracted the $\chi_{0l}$ spinors using \eref{SHF_chinorm}. 
The crucial identities for this calculation are 
\bea
(\hat p_1 \hat \epsilon_0^+)|_{z=z_1} & =&  
{ \la y p_1 0 \rb   \over \epsilon \sqrt 2  \la 0y \ra}  
= {1 \over \epsilon} ( p_1 \epsilon_0^+)|_{\zeta = y} ,  \quad (\hat p_2 \hat \epsilon_0^+)|_{z=z_2}  =  
{  \la y p_2 0 \rb   \over \epsilon \sqrt 2   \la 0y \ra}  
= {1 \over \epsilon} ( p_2 \epsilon_0^+)|_{\zeta = y} , 
\nnl 
(\hat p_2  \epsilon_3^-)|_{z=z_1} & =&  
{   \la 3 p_2 0 \rb  \over  \sqrt 2  \lb 3 0 \rb}  
=  ( p_2 \epsilon_3^-)|_{\tilde \zeta = \tilde \lambda_0} , \qquad (\hat p_1 \epsilon_3^-)|_{z=z_2} =  
{ \la 3 p_1 0 \rb  \over  \sqrt 2   \lb 3 0 \rb }  
= ( p_1 \epsilon_3^-)|_{\tilde \zeta = \tilde \lambda_0} . 
\eea 
One should stress that these are valid for any reference spinors $\zeta$ and $\tilde \zeta$; no ``gauge-fixing" was performed along the way. 
Then  Eq.~(\ref{eq:ST_exp_1}) can be written as 
\bea
\label{eq:ST_exp_2}
\hspace{-2cm} \hat \cM^{0,\epsilon}(\mathbf{1}_X \mathbf{2}_{\bar X} 3_\gamma^- 0_\gamma^+) & = &   
{(-1)^{2S} 2 q^2 e^2 \la 3 p_1 0 \rb \over \epsilon^2 m^{2S}  \lb 3 0\rb
\la 0 y\ra } \bigg \{  
- { \la y p_1 0 \rb \over u - m^2} 
 \exp \left ( - i { \epsilon_{3}^{- \, \mu} p_3^\nu J_2^{\mu\nu}  \over (p_2 \epsilon_3^-)|_{\tilde \zeta = \tilde \lambda_0}} \right )   \la \2\1 \ra^{2S}
\nnl & + & 
{\la y p_2 0 \rb \over t - m^2} 
   \exp \left (-  i { \epsilon_{3}^{- \, \mu} p_3^\nu J_1^{\mu\nu}  \over (p_1 \epsilon_3^-)|_{\tilde \zeta = \tilde \lambda_0}} \right )    \la \2\1 \ra^{2S} 
 \bigg \} + B_\infty .  
\eea 
In the exponentials, the numerators do not depend on the reference spinors thanks to the antisymmetry of $J^{\mu\nu}$. On the other hand, in the denominators,  $(p_i \epsilon_3^-)$ do depend on the reference spinors. 
Our calculation unambiguously dictates that these have to be evaluated at a particular reference spinor, 
$\tilde \zeta = \tilde \lambda_0$.\footnote{%
This iluminates one tiny detail in the derivation of  Ref.~\cite{Guevara:2018wpp}, where the choice of the reference spinor was made ad-hoc.}  
The two exponentials are in fact equal, 
so we can pull them in front of the curly bracket. Then,  summing the $t$- and $u$-poles, the dependence on the shift spinor $y$ cancels out and we obtain the formula for the Compton amplitude in the exponential form~\cite{Arkani-Hamed:2019ymq}: 
\beq
 \hat \cM^{0,\epsilon}(\mathbf{1}_X \mathbf{2}_{\bar X} 3_\gamma^- 0_\gamma^+) = 
(-1)^{2S} { 2 q^2 e^2 \la 3 p_1 0 \rb^2 \over 
 \epsilon^2 m^{2S}  (t-m^2) (u-m^2) } 
 \exp \left (-  i { \epsilon_{3}^{- \, \mu} p_3^\nu J_1^{\mu\nu}  \over (p_1 \epsilon_3^-)|_{\tilde \zeta = \tilde \lambda_0}} \right )    \la \2\1 \ra^{2S}  + B_\infty.  
\eeq 
This proves exponentiation of IR terms in Compton scattering for any matter spin. 
Recall that $B_\infty$ starts at $\cO(\epsilon^1)$, and thus qualifies as a UV term.  
Given its explicit form in \eref{QED_comptonS_UVeven}, it does not appear to have a simple exponential representation. 
Finally, notice that the exponential can be rewritten in a more explicit form as  
\beq
 \hat \cM^{0,\epsilon}(\mathbf{1}_X \mathbf{2}_{\bar X} 3_\gamma^- 0_\gamma^+) = 
 (-1)^{2S} { 2 q^2 e^2 \la 3 p_1 0 \rb^2  \over 
 \epsilon^2 m^{2S}  (t-m^2) (u-m^2) } 
 \exp \left ( {\lb 3 0\rb   \la 3 \1\ra (\lambda_3 \partial_{\chi_1})  \over
  \la 3 p_1 0 \rb  } \right )    \la \2\1 \ra^{2S}   + B_\infty . 
\eeq

\subsection{Soft exponentiation}

 Ref.~\cite{He:2014bga} showed that the soft theorem for gravitons and gluons could be written in terms of an exponential operator acting on lower point amplitudes. However, the proof is only valid for the minimal-helicity-violating (MHV) sector, as it relies on the  vanishing of the boundary terms discussed in~\cite{Cachazo:2014fwa}. 
 Using the soft recursion we can show that the exponential form is actually valid beyond the MHV sector. In fact, the exponential operator can be extended for amplitudes with massive particles of any spin and with a soft particle of any integer helicity $|h| \leq 2$. 
We begin by rewriting the minimal 3-point amplitude for $h \geq 0$ as 
 \bea
\label{eq:QED_Mffh_massive_minimal} 
\cM(\mathbf{1}_X \mathbf{2}_{\bar{X}}3^{-h}) = g_X  \left({\la 3 p_1 \zeta  \rb \over \la 3\zeta \ra}\right)^h {\lb \2\1 \rb^{2S} \over m^{2S-1+h} } 
,   \qquad   
 \cM(\mathbf{1}_X \mathbf{2}_{\bar{X}}3^h)  =    g_X   \left({\la \zeta p_1 3 \rb \over \lb 3\zeta \rb}\right)^h {\la \2\1 \ra^{2S} \over m^{2S-1+h} }    ,
\eea 
where $g_X = \sqrt 2 e q$ for $|h| = 1$, 
and $g_X = m/\mpl$ for $|h|=2$.   
The soft expansion in  Eq.~(\ref{eq:ST_softGravity})
and Eq.~(\ref{eq:ST_photonRaw}) can be collectively written as:
\beq 
\label{eq:ST_softExp}
\hat \cM^{0,\epsilon}(1 \dots n 0^+) =  {1 \over  \epsilon^{1+h} \,m^{1+h} } \sum_{l = 1}^n  
 { g_l \la y p_l 0 \rb^h  \over (2 p_0 p_l) \la 0 y \ra^h  }  
  \hat \cM^{z_l,\epsilon}(1 \dots P_{0l} \dots  n)
+ \cO(\epsilon^0). 
\eeq 
Now, the Taylor expansion in Eq.~(\ref{eq:ST_epsilonExpansion})
can be (trivially) recast in the exponential form: 
\beq 
\hat \cM^{z_l,\epsilon}(1 \dots P_{0l} \dots  n)   = \exp\left(\epsilon   { \la 0 y \ra \over \la y p_l 0 \rb } \tilde  {\cal D }_l  \right)
 \hat \cM^{0,0}(1 \dots  n) +  \cO(\epsilon^3) .
\eeq 
This way, for minimal coupling,  we obtain a similar form of the soft theorem as the one in  Ref.~\cite{He:2014bga}:
\beq 
\label{eq:ST_softExp2}
 \cM^{0,\epsilon}(1 \dots n 0^+) =  {1 \over  \epsilon^{1+h} \,m^{1+h}  } \sum_{l = 1}^n 
 { g_X \la y p_l 0 \rb^h  \over (2 p_0 p_l) \la 0 y \ra^h  }   \exp\left(\epsilon   { \la 0 y \ra \over \la y p_l 0 \rb } \tilde  {\cal D }_l  \right)
 \hat \cM^{0,0}(1 \dots  n) 
+ \cO(\epsilon^0). 
\eeq 
where the differential operator 
$\tilde  {\cal D }_l$ is given in Eq.~\eqref{eq:ST_calDl}.
This form may be useful to study amplitudes in the Mellin space along the lines of Ref.~\cite{Guevara:2019ypd}.
Moreover, our results in Eq.~\eqref{eq:subleadingSF_dipole} for the electromagnetic  dipole suggest that a generalization (for soft photons) consists in  replacing
\beq
\tilde  {\cal D }_l \rightarrow \tilde  {\cal D }_l 
 +   {S a_l \over q_l m }  
 [0 \mathbf{l}]^2  ,
\eeq
where for massive $l$ the suppressed  little group indices can be read off from Eq.~\eqref{eq:subleadingSF_dipole}. 
Similarly, for gravitons,
\eref{subsubleadingSFgravity_quadrupole} can be reproduced via 
$\tilde  {\cal D }_l \rightarrow \tilde  {\cal D }_l  
+ \epsilon  Q_l {\la y p_l 0 \rb  \over 2 \la 0 y \ra}  [0 \mathbf{l}]^4$.

\section{Conclusions}
\label{sec:conc}
\setcounter{equation}{0}

This paper provides a new toolbox to study theories with particles of arbitrary mass and spin. 
We introduced a class of complex momentum shifts combining a BCFW-like spinor shift with a soft limit of a massless particle present in the process.  
The important technical novelty is that our shifts are compatible with the massive spinors introduced by Arkani-Hamed, Huang, and Huang in Ref.~\cite{Arkani-Hamed:2017jhn}.
Via the usual road of the Cauchy theorem, 
these  shifts allow one to derive novel recursion relations for scattering amplitudes, even when all but one particles involved are massive. 

This formal development will find multiple practical applications, we hope.    
In this paper we applied it to derive a new incarnation of the soft theorems for emission of a soft photon or a graviton. 
They function, out of the box, with amplitudes expressed in the on-shell spinor formalism, also with massive spinors, thus generalizing the massless results of Ref.~\cite{Elvang:2016qvq}.
This allows us to systematically explore soft theorems for general theories,  and make a direct connection with the multipole expansion of the on-shell 3-point interaction between matter and photons/gravitons. 
In particular, our soft theorems lead to an elegant proof that gravitational dipole interactions are incompatible with the assumption of locality, unitarity and Poincar\'{e} invariance. 
We also provide a concise expression for the modification of the soft theorems due to the quadrupole (for gravitons) and dipole (for photons) non-minimal interactions with matter. 

Generally, soft theorems only capture the leading terms in the limit where the massless particle momentum goes to zero, because they take into account only the soft particle emission from external legs.
However, for Compton scattering, emission from external legs accounts for all factorization channels. 
Therefore, Compton amplitudes derived via soft theorems reflect the complete structure of physical poles required by unitarity.  
A subtlety appears for spin $S \geq 3/2$ (for photons) and $S \geq 5/2$ (for gravitons) when these amplitudes develop a spurious pole, which has to be subtracted by adding a judiciously designed UV term.  
Our approach leads to a new perspective on these UV terms: they arise as boundary terms in our recursion relations.   
We laid out an algorithm for constructing the UV term for Compton scattering of photons, 
which leads to a more compact and transparent formula than the one in Ref.~\cite{Chung:2018kqs}. 
We also derived the UV term for Compton scattering via the electromagnetic dipole.     
These results may be useful for practical applications, for example in nuclear or atomic physics.
Furthermore, our soft recursion leads by the shortest route to recasting the Compton amplitude into a form where  the increasing matter spin  corresponds to an action of an exponentiated angular momentum operator~\cite{Guevara:2018wpp,Arkani-Hamed:2019ymq}.  
We were also able to rewrite more general soft recursions in the exponentiated form, extending the results of Ref.~\cite{He:2014bga}.

Our soft recursions may help with calculation of multi-leg amplitudes - the topic we only fleetingly touched in this paper. 
That could in turn boost the amplitude methods of classical calculations of electromagnetic and gravitational radiation along the lines of Ref.~\cite{Kosower:2018adc}.  
These recursions should also be useful to derive more general soft theorems for Goldstone fermions~\cite{Chen:2014xoa} and gravitinos~\cite{Jain:2018fda}. 
Furthermore, it would  be interesting to generalize our recursions to multiple soft particles~\cite{Klose:2015xoa} (as recently explored with the CHY formalism \cite{Chakrabarti:2017zmh}).  
More speculatively, the tools we provide may open up new areas of exploration of massive amplitudes. 
One direction to mention in this context is the link with the asymptotic symmetries~\cite{Strominger:2017zoo,Law:2020tsg}, 
another is the double copy structure of the soft theorems for massive particles~\cite{Johansson:2019dnu}.

\section*{Acknowledgments}

AF is partially supported by the French Agence Nationale de la Recherche (ANR) under grant ANR-19-CE31-0012 (project MORA). The work of C.S.M.\ is supported by the Alexander von Humboldt Foundation, in the framework of the Sofja Kovalevskaja Award 2016, endowed by the German Federal Ministry of Education and Research and also supported by  the  Cluster  of  Excellence  ``Precision  Physics,  Fundamental Interactions, and Structure of Matter'' (PRISMA$^+$ EXC 2118/1) funded by the German Research Foundation (DFG) within the German Excellence Strategy (Project ID 39083149).

\appendix
\renewcommand{\theequation}{\Alph{section}.\arabic{equation}}

\section{Summary of spinor shifts}
\label{app:shifts}
\setcounter{equation}{0}

In this appendix we list distinct spinor shifts realizing the momentum shift in \eref{ST_ss}.  
We label them as $ \{ 0\bar A \bar B \}$ or $\{ \bar 0 A B\}$, where $A(B) = 0$ when the $j$-th ($k$-th) particle is massless, 
and $A(B) = m$ when the $j$-th ($k$-th) particle is massive. 
The label $\{ 0\bar A \bar B\}$ ($\{\bar 0 A B\}$) corresponds to shifting the holomorphic(antiholomorphic) spinor of the massless particle $0$, and  the antiholomorphic(holomorphic) spinors of the $j$-th and $k$-th particles. 

\subsection*{${\bf \{0\bar 0 \bar 0\}}$}

\beq
 q_0 \sigma = y \tilde \lambda_0, \quad 
q_j \sigma = \lambda_j \tilde \lambda_0, \quad 
q_k \sigma = \lambda_k  \tilde \lambda_0. 
\eeq
\bea
 \lambda_{0}^{z}  & = &  \epsilon  \lambda_0 - z  y, 
\nnl 
\tilde \lambda_j^z  & = & \tilde \lambda_j  -  
{ (\epsilon - 1) \langle 0k \rangle -  z  \langle y k\rangle \over \langle j k \rangle } \tilde  \lambda_0,  
 \nnl 
 \tilde \lambda_k^z  & = & \tilde \lambda_k   +
{ (\epsilon - 1) \langle 0j \rangle -  z  \langle y j\rangle \over \langle j k \rangle } \tilde  \lambda_0 .  
\eea

\subsection*{${\bf \{\bar 0 0  0\}}$} 

\beq
q_0 \sigma = \lambda_0 \tilde y, \quad 
q_j \sigma = \lambda_0 \tilde \lambda_j, \quad 
q_k \sigma = \lambda_0 \tilde \lambda_k . 
\eeq 
\bea
\label{eq:ST_000_AH-H}
\tilde \lambda_{0}^{z} & = &  \epsilon \tilde \lambda_0 - z \tilde y, 
\nnl 
\lambda_j^z  & = & \lambda_j  
-  { (\epsilon-1) [0k]  - z  [y k ] \over [jk]}   \lambda_0,  
 \nnl 
 \lambda_k^z  & = & \lambda_k   + {(\epsilon-1) [0j] - z  [y j]\over  [jk]}   \lambda_0. 
\eea

\subsection*{${\bf \{0\bar m \bar 0\}}$} 

\beq
 q_0 \sigma = y \tilde \lambda_0, \qquad 
q_j \sigma = p_j \sigma \tilde \lambda_0 \tilde \lambda_0, \qquad 
q_k \sigma = \lambda_k  \tilde \lambda_0 .
\eeq 
\bea
\lambda_{0}^{z}  & = &  \eps  \lambda_0 - z  y, 
\nnl 
\tilde \chi_j^z & = & \tilde \chi_j 
-  {  (\epsilon -1) \la k 0 \ra    - z  \la k y \ra  
\over  \la k p_j 0 \rb  } \lb \jb  \tilde 0 \rb \tilde \lambda_0 , 
 \nnl 
 \tilde \lambda_{k}^{z} & = & 
 \tilde \lambda_k   
 - { (\epsilon-1) \la 0 p_j 0 \rb- z \la y p_j 0  \rb  
 \over  \la k p_j 0 \rb} \tilde \lambda_0   . 
\eea

\subsection*{${\bf \{\bar 0 m  0\}}$} 

\beq
 q_0 \sigma = \lambda_0 \tilde y, \qquad 
q_j \sigma =  \lambda_0 \lambda_0 p_j \bar \sigma, \qquad 
q_k \sigma =   \lambda_0  \tilde  \lambda_k.
\eeq 
\bea
\tilde \lambda_{0}^{z} & = &  \epsilon \tilde \lambda_0 - z \tilde y, 
\nnl 
 \chi_j^z & = &  \chi_j 
-  { (\epsilon -1) [ k 0]    - z  [k y]
\over  \la 0  p_j k \rb  } [\mathbf{j} 0] \lambda_0 , 
 \nnl 
 \lambda_k^z & = & \lambda_k   
 - { (\epsilon-1) \langle 0 p_j 0 ] 
 - z \langle 0 p_j y ] 
 \over  \la  0 p_j k \rb} \lambda_0   . 
\eea
\subsection*{${\bf \{0\bar m \bar m\}}$} 

\beq
q_0 \sigma = y \tilde \lambda_0, \qquad 
q_j \sigma = p_j \sigma \tilde \lambda_0 \tilde \lambda_0, 
\qquad 
q_k \sigma = p_k \sigma \tilde \lambda_0 \tilde \lambda_0.  
\eeq
\bea
\lambda_{0}^z & = &  \eps  \lambda_0 - z  y, 
\nnl 
\tilde \chi_j^z & = & \tilde \chi_j  
+  { (\epsilon -1 )  \la  0 p_k 0 \rb    - z  \la y  p_k 0  \rb    \over  \lb 0 p_j p_k 0\rb } \lb \jb 0 \rb  \tilde \lambda_0 , 
 \nnl 
 \tilde \chi_k^z & = &
\tilde \chi_k   
 -  {  (\epsilon -1 )  \la 0 p_j 0 \rb  - z  \la y p_j 0  \rb    \over  \lb 0 p_j p_k 0\rb }  \lb \mathbf{k} 0 \rb \tilde \lambda_0 . 
\eea

\subsection*{${\bf \{ \bar 0 m  m \}}$} 

\beq
\label{eq:ST_0barmm_qsigma}
q_0 \sigma =  \lambda_0  \tilde y, \qquad 
q_j \sigma =  \lambda_0 p_j \bar \sigma  \lambda_0 , 
\qquad 
q_k \sigma =  \lambda_0 p_k \bar \sigma  \lambda_0 .
\eeq
\bea
\label{eq:ST_0barmm_spinor}
\tilde \lambda_{0}^{z}  & = &  \epsilon   \tilde \lambda_0 - z  \tilde  y, 
\nnl 
\chi_j^z & = & \chi_j 
+  {  (\epsilon -1) \la  0 p_k 0 \rb     - z  \la  0 p_k y  \rb 
\over  \la 0 p_j p_k 0 \ra  }    \la \jb 0 \ra   \lambda_0 , 
 \nnl 
 \chi_k^z & = & \chi_k
-  {  (\epsilon -1) \la 0 p_j 0  \rb    - z  \la  0 p_j y \rb  
\over \la 0 p_j p_k 0 \ra  }   \la \kb 0 \rangle   \lambda_0 .
\eea

\section{UV terms in Compton scattering}
\label{app:uvCompton}
\setcounter{equation}{0}

We have seen in \sref{app} that, when spin $S$ of the matter particle is larger than one,  the IR part of the Compton amplitude  obtained via soft theorems develops an unphysical pole at $\langle 3 p_1  4] \to 0$. 
The same happens for $S>2$ in gravitational Compton scattering. 
The unphysical pole has to be canceled by the boundary term in the recursion relation, which we refer to as the {\em UV term} in the discussion below.   
In this appendix we present an algorithm to construct a correct UV term, alternative to the one in Ref.~\cite{Chung:2018kqs}. 

For this discussion we find it convenient to introduce the auxiliary objects: 
\bea
\label{eq:UV_xyz}
{\cal A} &  \equiv &  \langle \mathbf{1} 3 \rangle [\mathbf{2} 4] ,
\qquad  
{\cal B}   \equiv   \langle \mathbf{2} 3 \rangle [\mathbf{1} 4],
\qquad 
{\cal X }    \equiv    
m \langle \mathbf{12}  \rangle + m [\mathbf{12}]
-\langle \mathbf{1} p_4 \mathbf{2}] -\langle \mathbf{2} p_3 \mathbf{1}] 
\nnl 
{\cal Y} & \equiv & (u-m^2) {\cal A} 
+ (t-m^2) {\cal B}  , 
\qquad 
{\cal Z}  \equiv  (u-m^2) {\cal A} 
- (t-m^2)  {\cal B} . 
\eea 
They satisfy the important identities 
\beq
\label{eq:UV_id}
\langle 3 p_1 4] \mathcal{X} -\mathcal{Y} =  
2 m^2 \left ({\cal A} + {\cal B} \right ),
\qquad 
\mathcal{Z}  = -\langle 3 p_1 4] 
\left (\langle \mathbf{1} p_4 \mathbf{2}] +\langle \mathbf{2} p_4 \mathbf{1}] \right ). 
\eeq 

\subsection{Electromagnetic Compton scattering}
\label{app:uvComptonEM}

The opposite helicity Compton scattering amplitude for matter particle $X$ of mass $m$ and spin $S$ can be written as
$\cM(\mathbf{1}_X \mathbf{2}_{\bar X} 3_\gamma^- 4_\gamma^+) =  \cM_{\rm IR} + \cM_{\rm UV}$.  
The IR term takes the form~\cite{Arkani-Hamed:2017jhn}:
\beq 
\label{eq:UV_McomptonIR}
 \cM_{\rm IR} = (-1)^{2S} 
{2 e^2  \langle 3 p_1  4]^2 \over (t-m^2) (u-m^2)} 
 \left [
 { {\cal A} + {\cal B} \over  
 \langle 3 p_1  4] } \right ]^{2S} , 
\eeq 
where $t = (p_1 + p_3)^2$, $u = (p_1 + p_4)^2$.  
As discussed in~\cite{Arkani-Hamed:2017jhn}, the IR term fully accounts for the kinematic $t$- and $u$-channel poles required by unitarity. 
The UV term can be zero or a pure contact term for $S \leq 1$. 
However, for $S> 1$ the IR term develops an unphysical pole at $\langle 3 p_1  4] \to 0$. 
Therefore the UV term must be necessarily non-zero to cancel the unphysical pole, without screwing up the physical $t$- and $u$-poles. 

To understand the general formula for arbitrary spin $S$, 
it is useful start with a simpler example of $S=2$. 
The first identity in \eref{UV_id} allows one to reshuffle the IR term, so as to rewrite the residue of the unphysical pole in a more convenient form~\cite{Chung:2018kqs}.
Plugging it into \eref{UV_McomptonIR}, the IR term for $S=2$ can be rewritten as  
\bea 
\label{eq:UV_MirS2}
\cM_{\rm IR} &= & 
{2 e^2  \langle 3 p_1  4]^2 \over (2m^2)^4 (t-m^2) (u-m^2)} 
 \left [
 { \langle 3 p_1 4] \mathcal{X} -\mathcal{Y}  \over  
 \langle 3 p_1  4] } \right ]^{4} 
 \nnl & = &  
 {2 e^2   \over (2m^2)^4 (t-m^2) (u-m^2)} 
 \left [ {\mathcal{Y}^4 \over  \langle 3 p_1  4]^2}
 - {4  \mathcal{X} \mathcal{Y}^3 
 \over  \langle 3 p_1  4]}  + \dots \right ] , 
\eea 
where the dots stand for terms with no unphysical poles.
One can cancel the double pole in $ \langle 3 p_1  4] $ by
including in the $\cM_{\rm UV}$ the term 
$- {2 e^2   \over (2m^2)^4 (t-m^2) (u-m^2)} 
{\mathcal{Y}^2 (\mathcal{Y}^2 - \mathcal{Z}^2 ) \over  \langle 3 p_1  4]^2}$. 
This would however tamper with the $t$- and $u$-poles, which we want to avoid. 
The trick is to compensate powers of ${\cal Y}$ with powers of ${\cal Z}$ which, by the second identity in \eref{UV_id}, do not affect the residue at $\langle 3 p_1  4] \to 0$. 
The crucial observation is that 
\beq
\frac{\mathcal{Y}^2-\mathcal{Z}^2}{(t-m^2) (u-m^2)}
= 4  {\cal A}   {\cal B} 
\eeq
does not have a pole at $t \to m^2$ or $u \to m^2$. 
Therefore, a UV term proportional to $\frac{\mathcal{Y}^2-\mathcal{Z}^2}{(t-m^2) (u-m^2)}$ will affect the $\langle 3 p_1  4]$ poles without destroying the correct structure of physical poles provided by the IR term. 
In our case,  the double pole at $\langle 3 p_1  4] \to 0$  is canceled if the UV term contains 
$\cM_{\rm UV} \supset - {2 e^2   \over (2m^2)^4 (t-m^2) (u-m^2)} 
{\mathcal{Y}^4 - \mathcal{Z}^4  \over  \langle 3 p_1  4]^2}$. 
In exactly the same way we can deal with the single 
$\langle 3 p_1  4]$ pole in \eref{UV_MirS2}. 
All in all, a legal UV term canceling all $\langle 3 p_1  4]$ poles in the $S=2$ Compton amplitude is given by 
\bea 
\label{eq:UV_MuvS2_1}
\cM_{\rm UV} & = &  
- {2 e^2   \over (2m^2)^4 (t-m^2) (u-m^2)} 
\bigg \{    {\mathcal{Y}^2 \big ( \mathcal{Y}^2 - \mathcal{Z}^2 \big)   \over  \langle 3 p_1  4]^2}
- {4  \mathcal{X} \mathcal{Y} (\mathcal{Y}^2 - \mathcal{Z}^2)
 \over  \langle 3 p_1  4]} 
 \bigg  \} 
 \nnl  &= &  - { e^2 {\cal A}   {\cal B}   \over 2m^8} 
\left \{ {\mathcal{Y}^2   \over  \langle 3 p_1  4]^2}
- {4  \mathcal{X} \mathcal{Y}
 \over  \langle 3 p_1  4]} \right \}  + C, 
\eea
where $C$ is any pure contact term. 
This construction guarantees that $\cM_{\rm IR} + \cM_{\rm UV}$ has no unphysical poles, and that it has the same  physical poles as $\cM_{\rm IR}$. 
The UV term is ultraviolet in two ways. 
One,  it scales as $\cO(\epsilon^1)$ in the (complex kinematics) limit where 
$t \to m^2 $, $u \to m^2$, and $s \to 0$ which is the limit where the photons become soft. 
This is unlike the IR term which scales as $\cO(\epsilon^{-2})$ in the soft limit.
Two, $\cM_{\rm UV}$ grows for $E \gg m$ as $\cO(E^8/m^8)$, unlike the IR term, which is regular at high energies. 
This growth leads to a loss of perturbative unitarity at the scale $\Lambda$ close to the mass $m$ of the particle $X$, which is what ultimately prevents the possibility of fundamental spin-2 massive particles.  
In fact, we can further massage \eref{UV_MuvS2_1} so as to improve the UV behavior by one notch, so as to (slightly) delay the onset of strong coupling. 
Using the first identity in \eref{UV_id} to eliminate ${\cal Y}$, and absorbing the resulting ${\cal X}^2$ terms into $C$, we end up with 
\beq
\label{eq:UV_MuvS2_2}
\cM_{\rm UV} =   
  - {2 e^2{\cal A}   {\cal B}   \over m^6} 
\left \{ { 
m^2 \left ( {\cal A}  +  {\cal B}  \right )^2   \over  \langle 3 p_1  4]^2}
+{\mathcal{X} \left ( {\cal A}  + {\cal B}   \right )
 \over  \langle 3 p_1  4]} \right \} 
+  C'  . 
\eeq
This has a better UV behavior, $\cM_{\rm UV} \sim \cO(E^6/m^6)$. 
Since the UV terms in \eref{UV_MuvS2_1} and  \eref{UV_MuvS2_2} differ only by a contact term, they have exactly  the same pole structure, and they both cancel the unphysical pole in $\cM_{\rm IR}$. 

Following exactly the same steps one can construct $\cM_{\rm UV}$ for arbitrary $S > 1$: 
\beq
\label{eq:UV_Muv}
\cM_{\rm UV}  =   
 -  { 2 e^2  \over  (2 m^2)^{2S} (t-m^2) (u-m^2)}  
 \sum_{k= 3}^{2S} 
 \binom{2S}{k}  {  (-{\cal X})^{2S - k}  {\cal Y} \left [ {\cal Y}^{k-1}  -  \{{\cal Y} || {\cal Z} \} {\cal Z}^{k-2}  \right ] \over 
\langle 3 p_1  4]^{k-2} }, 
\eeq 
where
$\{{\cal Y} || {\cal Z} \} = {\cal Y} ({\cal Z})$ for even (odd) $k$. 
Note that ${\cal X}$ and  ${\cal Z}$ are odd under $1 \leftrightarrow 2$, while $ {\cal Y}$ is even.  
Therefore $\cM_{\rm UV}$ has the same symmetry under 
$1 \leftrightarrow 2$ as the IR term: 
it is even (odd) for integer (half-integer) spin.  
The absence of physical poles in $\cM_{\rm UV}$ follows from 
${\cal Y}^{n} - {\cal Z}^{n}= ({\cal Y}^{2} - {\cal Z}^{2})
({\cal Y}^{n-2} +{\cal Y}^{n-4}  {\cal Z}^2 + \dots  +  {\cal Z}^{n-2 })$ 
for any even $n \geq 2$.
This is the most compact form of the UV term, 
but not the one with the best possible UV behavior.
As in the case of $S=2$, we can always soften the UV behavior 
from $\cO(E^{4S}/m^{4S})$ down to $\cO(E^{4S-2}/m^{4S-2})$ by using the first identity in \eref{UV_id} and isolating the resulting $\cO(E^{4S}/m^{4S})$ contact term. 
The consequence is that the maximum cutoff scale of an effective theory describing a spin-$S>1$ particle with mass $m$ and electric charge $q$ satisfies 
\beq
\Lambda_{\rm max}  \lesssim   \bigg (  {4 \pi \over |q| e}  \bigg  )^{1 \over 2S -1} . 
\eeq 
This bound was obtained by other methods in Ref.~\cite{Porrati:2008ha}.

\vspace{1cm}

For Compton scattering via the electromagnetic dipole the IR amplitude can be read off from 
\eref{APP_comptonDipole}:
\beq  
 \label{eq:UV_comptonDipole}
\cM_{\rm IR}   =   \displaystyle
  |a|^2 {(-1)^{2S} S e^2  \over 2 m^2 (t-m^2) (u-m^2)} 
 \left [
 { {\cal A}  +  {\cal B}   \over  
 \langle 3 p_1 4]  } \right ]^{2S-2} 
 \bigg \{
  ( {\cal Y } + {\cal Z}) 
\big(    {\cal A}    +2 S  {\cal B}  \big  )  
+ ( {\cal Y } - {\cal Z}) 
\big  (  2 S  {\cal A }  +  {\cal B} \big ) 
 \bigg  \}  . \qquad  \eeq 
The corresponding UV term can be easily constructed by the same techniques:    
\bea
\label{eq:UV_dipoleBoundary} 
\cM_{\rm UV}  & = & 
- |a|^2 { S e^2  \over  (2 m^2)^{2S-1} (t-m^2) (u-m^2) } 
\sum_{k = 1}^{2S-2} \binom{2S-2}{k}   {(-X)^{2S - 2 - k} \over   \langle 3 p_1 4]^k}
\bigg \{  
\nnl & & 
({\cal Y }^k  -  {\cal Z }^k) ( {\cal Y } + {\cal Z}) 
\big(  {\cal A}    +2 S  {\cal B}    \big  )  
+ 
({\cal Y }^k  -  (- {\cal Z })^k) ( {\cal Y } - {\cal Z}) 
\big  ( 2 S  {\cal A }  +  {\cal B}   \big ) 
\bigg \}. 
\nnl  
  \eea 
  
\subsection{Gravitational Compton scattering}
\label{app:uvComptonGR} 

We turn to the analogue of the Compton process in gravity, where photons are replaced by massless spin-2 gravitons. 
We assume minimal gravitational interactions of a spin-S massive particle $X$. 
We focus  on the opposite helicity Compton amplitude and again separate the UV and IR terms: 
$\cM(\mathbf{1}_X \mathbf{2}_X 3_h^- 4_h^+) =  \cM_{\rm IR} + \cM_{\rm UV}$.  
The IR term takes the form~\cite{Arkani-Hamed:2017jhn}:
\bea 
\label{eq:UV_MirGR}
\cM_{\rm IR}  & =&  
 {\big (  {\cal A }  +  {\cal B}    \big )^4 \over  \mpl^2 s (t-m^2)(u - m^2)}
\left [  {\cal A }  +  {\cal B}     \over  \langle 3 p_1  4]  \right ]^{2S-4} 
\nnl  &= &  {  {\cal A }  +  {\cal B}     \over \mpl^2 (2 m^2)^{2S-2}} 
\sum_{k=0}^{2S-3}  (-1)^k \binom{2S-3}{k} X^{2S -3 -k} I_k ,  
\eea 
where we abbreviated
\beq
I_k \equiv {2 m^2 \big (  {\cal A }  +  {\cal B}  \big )^2  \over s (t - m^2) (u-m^2)} {{\cal Y}^k \over \langle 3 p_1  4]^{k-1}},  
\eeq
and  ${\cal A}$, ${\cal B}$, ${\cal X}$, ${\cal Y}$ are defined in \eref{UV_xyz}. 
For $S>2$  the IR term in \eref{UV_MirGR} develops a  pole at $\langle 3 p_1  4]  \to 0$, therefore  $\cM_{\rm UV}$ has to be non-zero for $S>2$ so as to cancel the unphysical pole. 
Below we construct a befitting UV term.  
We will first find a set of functions $U_k$ such that
 $U_k + I_k$ is regular at $\langle 3 p_1  4]  \to 0$ and has the same kinematic poles as $I_k$. 
Given such a set, a legal UV term for $S>2$ is simply 
\beq
\label{eq:GR_muv_spinS}
\cM_{\rm UV }  = {  {\cal A }  +  {\cal B}  \over \mpl^2 (2 m^2)^{2S-2}} 
\sum_{k=2}^{2S-3}  (-1)^k \binom{2S-3}{k} X^{2S -3 -k} U_k. 
\eeq   
We will construct $U_k$ recursively. 
In this endeavour the following identity is instrumental:
\beq 
\label{eq:GR_ApB_central}
 \big (   {\cal A }  +  {\cal B}    \big )^2  = 
  {1 \over m^2} \bigg \{ 
s   {\cal A }   {\cal B} 
+m  \langle 3 p_1  4]  \big (   {\cal A }  +  {\cal B}  \big ) 
\big (  \langle \mathbf{12} \rangle   +  [\mathbf{12} ]  \big ) 
- \langle 3 p_1  4]^2 \langle \mathbf{12} \rangle    [\mathbf{12} ]  \bigg \}  . 
\eeq  
This allows us to rewrite $I_k$ as 
\beq 
I_k = { 2 {\cal A} {\cal B}  \over  (t - m^2) (u-m^2)} {{\cal Y}^k \over \langle 3 p_1  4]^{k-1}}
+ c_k \big (  \langle \mathbf{12} \rangle   +  [\mathbf{12} ]  \big )  
- d_k \langle \mathbf{12} \rangle    [\mathbf{12} ],
\eeq 
where once again we introduced some auxiliary notation: 
\beq
c_k \equiv 
 {2 m  \big ( {\cal A}  + {\cal B}  \big ) \over s (t - m^2) (u-m^2)} {{\cal Y}^k \over \langle 3 p_1  4]^{k-2}}, 
\qquad 
d_k \equiv {2  \over s (t - m^2) (u-m^2)} {{\cal Y}^k \over \langle 3 p_1  4]^{k-3}} . 
\eeq 
We can rearrange $c_k$ as 
\beq
c_k = 
 {2 m  \big ( {\cal A}  + {\cal B}  \big ) \over s (t - m^2) (u-m^2)} {{\cal Y}^{k-1} \over \langle 3 p_1  4]^{k-2}}
 \big [ -2 m ^2 \big ( {\cal A}  + {\cal B}  \big )  + {\cal X} \langle 3 p_1  4] \big ] 
 = -2 m I_{k-1} +   {\cal X}  c_{k-1} .  
\eeq 
This recursive equation is solved by 
\beq
c_k = - 2 m \sum_{l= 1}^{k-2} I_{k-l} {\cal X}^{l-1} +   {\cal X}^{k-2} c_2 . 
\eeq 
We can handle $d_k$ in a similar way 
\beq
d_k \equiv {2  \over s (t - m^2) (u-m^2)} {{\cal Y}^{k-2} \over \langle 3 p_1  4]^{k-3}}
 \big [ -2 m^2 \big ( {\cal A}  + {\cal B}  \big )  + {\cal X} \langle 3 p_1  4] \big ]^2 
 =  4 m^2 I_{k-2} - 4 m^2 {\cal X}  c_{k-2}  +   {\cal X}^2  d_{k-2} . 
\eeq 
The solution is 
\bea
d_k & = & 4 m^2 \sum_{l=1}^{(k-2)/2}\big (  I_{k-2 l}  -  {\cal X}  c_{k-2 l} \big ) {\cal X}^{2(l-1)}  
+  {\cal X}^{k-2}  d_{2},    \qquad   {\rm even} \ k , 
\nnl 
d_k & = & 4 m^2 \sum_{l=1}^{(k-3)/2}\big (  I_{k-2 l}  -  {\cal X}  c_{k-2 l} \big ) {\cal X}^{2(l-1)}  
+  {\cal X}^{k-3}  d_{3} ,   \qquad   {\rm odd} \ k . 
\eea 
Plugging the solution back into the  formula for $I_k$ we obtain 
\bea
I_k  & = &  
{ 2 {\cal A} {\cal B}  \over  (t - m^2) (u-m^2)} 
{{\cal Y}^k \over \langle 3 p_1  4]^{k-1}}
-2 m  \big (  \langle \mathbf{12} \rangle   +  [\mathbf{12} ]  \big )  \sum_{l= 1}^{k-2} I_{k-l} {\cal X}^{l-1} 
\nnl  && 
- 4m^2  \langle \mathbf{12} \rangle     [\mathbf{12} ]   
\sum_{l=1}^{(k-\{2||3 \})/2}\big (  I_{k-2 l}  +    
 2 m  {\cal X} \sum_{n= 1}^{k - 2l -2} I_{k-n} {\cal X}^{n-1}
 \big ) {\cal X}^{2(l-1)}  
+ \dots \qquad 
\eea  
where the dots stand for $d_{2,3}$ and $c_2$ terms, which do not have poles at $\langle 3 p_1  4]$. 
We are ready to write down a recursive equation for $U_k$: 
\bea
\label{eq:GR_uk_recursive}
U_k  & = &  
- { 2 {\cal A} {\cal B}  \over  (t - m^2) (u-m^2)} {{\cal Y}^k  - \{ {\cal Z}  || {\cal } Y \}   {\cal Z}^{k-1}\over \langle 3 p_1  4]^{k-1}}
-2 m  \big (  \langle \mathbf{12} \rangle   +  [\mathbf{12} ]  \big )   \sum_{l= 1}^{k-2} U_{k-l} {\cal X}^{l-1} 
\nnl  && 
- 4m^2   \langle \mathbf{12} \rangle    [\mathbf{12} ]  
\sum_{l=1}^{(k-\{2||3 \})/2}\big (  U_{k-2 l}  +    
 2 m  {\cal X} \sum_{n= 1}^{k - 2l -2} U_{k-n} {\cal X}^{n-1}
 \big ) {\cal X}^{2(l-1)}  , 
\qquad 
\eea 
where $ \{ {\cal Z}  || {\cal } Y \} = {\cal Z}  ({\cal Y} )$ for even (odd) $k$, and $\{2||3 \}  = 2(3)$  for even (odd) $l$. 
An implicit convention above is that, if the upper limit of a summation is smaller than the lower limit,  then the whole sum is void. 

The recursive equation~(\ref{eq:GR_uk_recursive}) supplemented by the boundary condition  $U_0 = U_1 = 0$ defines $U_k$ for arbitrary $k$.
Via \eref{GR_muv_spinS}, this allows one to construct an appropriate UV term for any spin $S$ of the matter particle. 
Let us see how it works in some simple examples.  
For $k=2$ we get 
\beq
U_2 =
- { 2 {\cal A} {\cal B}  \over  (t - m^2) (u-m^2)} 
{{\cal Y}^2  -  {\cal Z}^2 \over \langle 3 p_1  4]} 
= - { 8 {\cal A}^2 {\cal B}^2 \over \langle 3 p_1  4] } . 
\eeq 
By \eref{GR_muv_spinS}, the UV term for $S=5/2$ reads 
\beq
\label{eq:GR_muv_spin52}
{\bf S=5/2}: \qquad 
\cM_{\rm UV }  = { {\cal A}  + {\cal B}   \over \mpl^2 (2 m^2)^3 }   U_2
=   - { {\cal A}^2  {\cal B}^2  \big (  {\cal A}  + {\cal B}  \big ) \over \mpl^2  m^6  \langle 3 p_1  4]  } .
\eeq   
For $k = 3$ \eref{GR_uk_recursive} yields 
\beq
U_3 = - 8 {\cal A}^2 {\cal B}^2  \bigg ( {  {\cal Y} \over \langle 3 p_1  4]^2 }  
- {2 m \big (  \langle \mathbf{12} \rangle   +  [\mathbf{12} ]  \big ) \over \langle 3 p_1  4] }  \bigg )    . 
\eeq 
By \eref{GR_muv_spinS}, the UV term for $S=3$ reads  
\bea
\label{eq:GR_muv_spin3}
\cM_{\rm UV }  & =&  { {\cal A}  + {\cal B}   \over \mpl^2 (2 m^2)^4 } \bigg ( 3 {\cal X} U_2  - U_3 \bigg ) 
\nnl & =& 
-   { {\cal A}^2  {\cal B}^2  \big (  {\cal A}  + {\cal B}  \big ) \over  \mpl^2  m^8 } 
\bigg ( 
{ {\cal X} \over \langle 3 p_1  4] }  
+ { m  \big (  \langle \mathbf{12} \rangle   +  [\mathbf{12} ]  \big ) \over \langle 3 p_1  4] }  
 + { m^2  \big (  {\cal A}  + {\cal B}  \big ) \over \langle 3 p_1  4]^2 }   
\bigg ) . 
\eea    
And so on. 

From \eref{GR_uk_recursive} one can read off that  $U_k$ scales as  $\cO(E^{2 (k+1)})$ at large energies.  
For Compton processes with energies much above the particle's mass one thus finds that the UV term grows with energy as 
\beq
\label{eq:GR_muv_asymptotics}
\cM_{\rm UV}  \sim  \cO \left (E^{4S -2} \over  m^{4S-4} \mpl^2 \right ), \qquad E \gg m,  
\eeq 
for any $S>2$. 
This violates perturbative unitarity at some energy scale above the particle's mass. 
One concludes that the maximum cutoff scale  for an effective theory describing a particle of mass $m$ and spin $S>2$ satisfies  
\beq
\Lambda_{\rm max}  \lesssim   
\bigg (  4 \pi \mpl m^{2S-2}  \bigg  )^{1 \over 2S -1} . 
\eeq 
This bound was conjectured in Refs.~\cite{Rahman:2009zz,Bonifacio:2018aon}. 
The bound is totally general and universal, since every particle has to be coupled to gravity. 
As the spin increases the maximum cutoff quickly approaches $m$, squeezing the validity range of the effective theory to a very narrow energy interval.   
This imposes severe restrictions on constructing viable effective theories of higher-spin particles (see also Refs.~\cite{Caron-Huot:2016icg,Afkhami-Jeddi:2018apj} for other arguments leading to similar conclusions).

\bibliographystyle{JHEP} 
\bibliography{softmatters}

\providecommand{\href}[2]{#2}\begingroup\raggedright\begin{thebibliography}{10}

\bibitem{Low:1958sn}
F.~Low, {\it {Bremsstrahlung of very low-energy quanta in elementary particle
  collisions}},  {\em Phys. Rev.} {\bf 110} (1958) 974--977.

\bibitem{Weinberg:1965nx}
S.~Weinberg, {\it {Infrared photons and gravitons}},  {\em Phys. Rev.} {\bf
  140} (1965) B516--B524.

\bibitem{Broedel:2014fsa}
J.~Broedel, M.~de~Leeuw, J.~Plefka, and M.~Rosso, {\it {Constraining subleading
  soft gluon and graviton theorems}},  {\em Phys. Rev. D} {\bf 90} (2014),
  no.~6 065024, [\href{http://arxiv.org/abs/1406.6574}{{\tt arXiv:1406.6574}}].

\bibitem{Rodina:2018pcb}
L.~Rodina, {\it {Scattering Amplitudes from Soft Theorems and Infrared
  Behavior}},  {\em Phys. Rev. Lett.} {\bf 122} (2019), no.~7 071601,
  [\href{http://arxiv.org/abs/1807.09738}{{\tt arXiv:1807.09738}}].

\bibitem{Strominger:2017zoo}
A.~Strominger, {\it {Lectures on the Infrared Structure of Gravity and Gauge
  Theory}},  \href{http://arxiv.org/abs/1703.05448}{{\tt arXiv:1703.05448}}.

\bibitem{Cheung:2015ota}
C.~Cheung, K.~Kampf, J.~Novotny, C.-H. Shen, and J.~Trnka, {\it {On-Shell
  Recursion Relations for Effective Field Theories}},  {\em Phys. Rev. Lett.}
  {\bf 116} (2016), no.~4 041601, [\href{http://arxiv.org/abs/1509.03309}{{\tt
  arXiv:1509.03309}}].

\bibitem{Cheung:2016drk}
C.~Cheung, K.~Kampf, J.~Novotny, C.-H. Shen, and J.~Trnka, {\it {A Periodic
  Table of Effective Field Theories}},  {\em JHEP} {\bf 02} (2017) 020,
  [\href{http://arxiv.org/abs/1611.03137}{{\tt arXiv:1611.03137}}].

\bibitem{Cheung:2018oki}
C.~Cheung, K.~Kampf, J.~Novotny, C.-H. Shen, J.~Trnka, and C.~Wen, {\it {Vector
  Effective Field Theories from Soft Limits}},  {\em Phys. Rev. Lett.} {\bf
  120} (2018), no.~26 261602, [\href{http://arxiv.org/abs/1801.01496}{{\tt
  arXiv:1801.01496}}].

\bibitem{Elvang:2018dco}
H.~Elvang, M.~Hadjiantonis, C.~R.~T. Jones, and S.~Paranjape, {\it {Soft
  Bootstrap and Supersymmetry}},  \href{http://arxiv.org/abs/1806.06079}{{\tt
  arXiv:1806.06079}}.

\bibitem{Low:2019ynd}
I.~Low and Z.~Yin, {\it {Soft Bootstrap and Effective Field Theories}},  {\em
  JHEP} {\bf 11} (2019) 078, [\href{http://arxiv.org/abs/1904.12859}{{\tt
  arXiv:1904.12859}}].

\bibitem{Cachazo:2014fwa}
F.~Cachazo and A.~Strominger, {\it {Evidence for a New Soft Graviton Theorem}},
   \href{http://arxiv.org/abs/1404.4091}{{\tt arXiv:1404.4091}}.

\bibitem{Britto:2005fq}
R.~Britto, F.~Cachazo, B.~Feng, and E.~Witten, {\it {Direct proof of tree-level
  recursion relation in Yang-Mills theory}},  {\em Phys. Rev. Lett.} {\bf 94}
  (2005) 181602, [\href{http://arxiv.org/abs/hep-th/0501052}{{\tt
  hep-th/0501052}}].

\bibitem{Bern:1998sv}
Z.~Bern, L.~J. Dixon, M.~Perelstein, and J.~Rozowsky, {\it {Multileg one loop
  gravity amplitudes from gauge theory}},  {\em Nucl. Phys. B} {\bf 546} (1999)
  423--479, [\href{http://arxiv.org/abs/hep-th/9811140}{{\tt hep-th/9811140}}].

\bibitem{He:2014bga}
S.~He, Y.-t. Huang, and C.~Wen, {\it {Loop Corrections to Soft Theorems in
  Gauge Theories and Gravity}},  {\em JHEP} {\bf 12} (2014) 115,
  [\href{http://arxiv.org/abs/1405.1410}{{\tt arXiv:1405.1410}}].

\bibitem{Sahoo:2018lxl}
B.~Sahoo and A.~Sen, {\it {Classical and Quantum Results on Logarithmic Terms
  in the Soft Theorem in Four Dimensions}},  {\em JHEP} {\bf 02} (2019) 086,
  [\href{http://arxiv.org/abs/1808.03288}{{\tt arXiv:1808.03288}}].

\bibitem{Elvang:2016qvq}
H.~Elvang, C.~R.~T. Jones, and S.~G. Naculich, {\it {Soft Photon and Graviton
  Theorems in Effective Field Theory}},  {\em Phys. Rev. Lett.} {\bf 118}
  (2017), no.~23 231601, [\href{http://arxiv.org/abs/1611.07534}{{\tt
  arXiv:1611.07534}}].

\bibitem{Bianchi:2015lnw}
M.~Bianchi and A.~L. Guerrieri, {\it {On the soft limit of closed string
  amplitudes with massive states}},  {\em Nucl. Phys.} {\bf B905} (2016)
  188--216, [\href{http://arxiv.org/abs/1512.00803}{{\tt arXiv:1512.00803}}].

\bibitem{AtulBhatkar:2018kfi}
S.~Atul~Bhatkar and B.~Sahoo, {\it {Subleading Soft Theorem for arbitrary
  number of external soft photons and gravitons}},  {\em JHEP} {\bf 01} (2019)
  153, [\href{http://arxiv.org/abs/1809.01675}{{\tt arXiv:1809.01675}}].

\bibitem{Sen:2017nim}
A.~Sen, {\it {Subleading Soft Graviton Theorem for Loop Amplitudes}},  {\em
  JHEP} {\bf 11} (2017) 123, [\href{http://arxiv.org/abs/1703.00024}{{\tt
  arXiv:1703.00024}}].

\bibitem{Laddha:2017ygw}
A.~Laddha and A.~Sen, {\it {Sub-subleading Soft Graviton Theorem in Generic
  Theories of Quantum Gravity}},  {\em JHEP} {\bf 10} (2017) 065,
  [\href{http://arxiv.org/abs/1706.00759}{{\tt arXiv:1706.00759}}].

\bibitem{Chakrabarti:2017ltl}
S.~Chakrabarti, S.~P. Kashyap, B.~Sahoo, A.~Sen, and M.~Verma, {\it {Subleading
  Soft Theorem for Multiple Soft Gravitons}},  {\em JHEP} {\bf 12} (2017) 150,
  [\href{http://arxiv.org/abs/1707.06803}{{\tt arXiv:1707.06803}}].

\bibitem{Arkani-Hamed:2017jhn}
N.~Arkani-Hamed, T.-C. Huang, and Y.-t. Huang, {\it {Scattering Amplitudes For
  All Masses and Spins}},  \href{http://arxiv.org/abs/1709.04891}{{\tt
  arXiv:1709.04891}}.

\bibitem{Kosower:2018adc}
D.~A. Kosower, B.~Maybee, and D.~O'Connell, {\it {Amplitudes, Observables, and
  Classical Scattering}},  {\em JHEP} {\bf 02} (2019) 137,
  [\href{http://arxiv.org/abs/1811.10950}{{\tt arXiv:1811.10950}}].

\bibitem{Guevara:2018wpp}
A.~Guevara, A.~Ochirov, and J.~Vines, {\it {Scattering of Spinning Black Holes
  from Exponentiated Soft Factors}},  {\em JHEP} {\bf 09} (2019) 056,
  [\href{http://arxiv.org/abs/1812.06895}{{\tt arXiv:1812.06895}}].

\bibitem{Laddha:2018rle}
A.~Laddha and A.~Sen, {\it {Gravity Waves from Soft Theorem in General
  Dimensions}},  {\em JHEP} {\bf 09} (2018) 105,
  [\href{http://arxiv.org/abs/1801.07719}{{\tt arXiv:1801.07719}}].

\bibitem{Arkani-Hamed:2019ymq}
N.~Arkani-Hamed, Y.-t. Huang, and D.~O'Connell, {\it {Kerr black holes as
  elementary particles}},  {\em JHEP} {\bf 01} (2020) 046,
  [\href{http://arxiv.org/abs/1906.10100}{{\tt arXiv:1906.10100}}].

\bibitem{Bonifacio:2018vzv}
J.~Bonifacio and K.~Hinterbichler, {\it {Bounds on Amplitudes in Effective
  Theories with Massive Spinning Particles}},  {\em Phys. Rev.} {\bf D98}
  (2018), no.~4 045003, [\href{http://arxiv.org/abs/1804.08686}{{\tt
  arXiv:1804.08686}}].

\bibitem{Bonifacio:2018aon}
J.~Bonifacio and K.~Hinterbichler, {\it {Universal bound on the strong coupling
  scale of a gravitationally coupled massive spin-2 particle}},  {\em Phys.
  Rev. D} {\bf 98} (2018), no.~8 085006,
  [\href{http://arxiv.org/abs/1806.10607}{{\tt arXiv:1806.10607}}].

\bibitem{Bonifacio:2019mgk}
J.~Bonifacio, K.~Hinterbichler, and R.~A. Rosen, {\it {Constraints on a
  gravitational Higgs mechanism}},  {\em Phys. Rev.} {\bf D100} (2019), no.~8
  084017, [\href{http://arxiv.org/abs/1903.09643}{{\tt arXiv:1903.09643}}].

\bibitem{Falkowski:2020mjq}
A.~Falkowski and G.~Isabella, {\it {Matter coupling in massive gravity}},  {\em
  JHEP} {\bf 04} (2020) 014, [\href{http://arxiv.org/abs/2001.06800}{{\tt
  arXiv:2001.06800}}].

\bibitem{Shadmi:2018xan}
Y.~Shadmi and Y.~Weiss, {\it {Effective Field Theory Amplitudes the On-Shell
  Way: Scalar and Vector Couplings to Gluons}},
  \href{http://arxiv.org/abs/1809.09644}{{\tt arXiv:1809.09644}}.

\bibitem{Aoude:2019tzn}
R.~Aoude and C.~S. Machado, {\it {The Rise of SMEFT On-shell Amplitudes}},
  \href{http://arxiv.org/abs/1905.11433}{{\tt arXiv:1905.11433}}.

\bibitem{Durieux:2019eor}
G.~Durieux, T.~Kitahara, Y.~Shadmi, and Y.~Weiss, {\it {The electroweak
  effective field theory from on-shell amplitudes}},
  \href{http://arxiv.org/abs/1909.10551}{{\tt arXiv:1909.10551}}.

\bibitem{Chung:2018kqs}
M.-Z. Chung, Y.-T. Huang, J.-W. Kim, and S.~Lee, {\it {The simplest massive
  S-matrix: from minimal coupling to Black Holes}},  {\em JHEP} {\bf 04} (2019)
  156, [\href{http://arxiv.org/abs/1812.08752}{{\tt arXiv:1812.08752}}].

\bibitem{Guevara:2019fsj}
A.~Guevara, A.~Ochirov, and J.~Vines, {\it {Black-hole scattering with general
  spin directions from minimal-coupling amplitudes}},  {\em Phys. Rev. D} {\bf
  100} (2019), no.~10 104024, [\href{http://arxiv.org/abs/1906.10071}{{\tt
  arXiv:1906.10071}}].

\bibitem{Cheung:2016iub}
C.~Cheung, A.~de~la Fuente, and R.~Sundrum, {\it {4D scattering amplitudes and
  asymptotic symmetries from 2D CFT}},  {\em JHEP} {\bf 01} (2017) 112,
  [\href{http://arxiv.org/abs/1609.00732}{{\tt arXiv:1609.00732}}].

\bibitem{Pasterski:2017kqt}
S.~Pasterski and S.-H. Shao, {\it {Conformal basis for flat space amplitudes}},
   {\em Phys. Rev. D} {\bf 96} (2017), no.~6 065022,
  [\href{http://arxiv.org/abs/1705.01027}{{\tt arXiv:1705.01027}}].

\bibitem{Donnay:2018neh}
L.~Donnay, A.~Puhm, and A.~Strominger, {\it {Conformally Soft Photons and
  Gravitons}},  {\em JHEP} {\bf 01} (2019) 184,
  [\href{http://arxiv.org/abs/1810.05219}{{\tt arXiv:1810.05219}}].

\bibitem{Dreiner:2008tw}
H.~K. Dreiner, H.~E. Haber, and S.~P. Martin, {\it {Two-component spinor
  techniques and Feynman rules for quantum field theory and supersymmetry}},
  {\em Phys. Rept.} {\bf 494} (2010) 1--196,
  [\href{http://arxiv.org/abs/0812.1594}{{\tt arXiv:0812.1594}}].

\bibitem{Dittmaier:1998nn}
S.~Dittmaier, {\it {Weyl-van der Waerden formalism for helicity amplitudes of
  massive particles}},  {\em Phys. Rev.} {\bf D59} (1998) 016007,
  [\href{http://arxiv.org/abs/hep-ph/9805445}{{\tt hep-ph/9805445}}].

\bibitem{Conde:2016vxs}
E.~Conde and A.~Marzolla, {\it {Lorentz Constraints on Massive Three-Point
  Amplitudes}},  {\em JHEP} {\bf 09} (2016) 041,
  [\href{http://arxiv.org/abs/1601.08113}{{\tt arXiv:1601.08113}}].

\bibitem{Franken:2019wqr}
R.~Franken and C.~Schwinn, {\it {On-shell constructibility of Born amplitudes
  in spontaneously broken gauge theories}},  {\em JHEP} {\bf 02} (2020) 073,
  [\href{http://arxiv.org/abs/1910.13407}{{\tt arXiv:1910.13407}}].

\bibitem{Weinberg:1964ew}
S.~Weinberg, {\it {Photons and Gravitons in s Matrix Theory: Derivation of
  Charge Conservation and Equality of Gravitational and Inertial Mass}},  {\em
  Phys. Rev.} {\bf 135} (1964) B1049--B1056.

\bibitem{Witten:2003nn}
E.~Witten, {\it {Perturbative gauge theory as a string theory in twistor
  space}},  {\em Commun. Math. Phys.} {\bf 252} (2004) 189--258,
  [\href{http://arxiv.org/abs/hep-th/0312171}{{\tt hep-th/0312171}}].

\bibitem{Hansen:1974zz}
R.~Hansen, {\it {Multipole moments of stationary space-times}},  {\em J. Math.
  Phys.} {\bf 15} (1974) 46--52.

\bibitem{Guevara:2019ypd}
A.~Guevara, {\it {Notes on Conformal Soft Theorems and Recursion Relations in
  Gravity}},  \href{http://arxiv.org/abs/1906.07810}{{\tt arXiv:1906.07810}}.

\bibitem{Chen:2014xoa}
W.-M. Chen, Y.-t. Huang, and C.~Wen, {\it {New Fermionic Soft Theorems for
  Supergravity Amplitudes}},  {\em Phys. Rev. Lett.} {\bf 115} (2015), no.~2
  021603, [\href{http://arxiv.org/abs/1412.1809}{{\tt arXiv:1412.1809}}].

\bibitem{Jain:2018fda}
D.~Jain and A.~Rudra, {\it {Leading soft theorem for multiple gravitini}},
  {\em JHEP} {\bf 06} (2019) 004, [\href{http://arxiv.org/abs/1811.01804}{{\tt
  arXiv:1811.01804}}].

\bibitem{Klose:2015xoa}
T.~Klose, T.~McLoughlin, D.~Nandan, J.~Plefka, and G.~Travaglini, {\it
  {Double-Soft Limits of Gluons and Gravitons}},  {\em JHEP} {\bf 07} (2015)
  135, [\href{http://arxiv.org/abs/1504.05558}{{\tt arXiv:1504.05558}}].

\bibitem{Chakrabarti:2017zmh}
S.~Chakrabarti, S.~P. Kashyap, B.~Sahoo, A.~Sen, and M.~Verma, {\it {Testing
  Subleading Multiple Soft Graviton Theorem for CHY Prescription}},  {\em JHEP}
  {\bf 01} (2018) 090, [\href{http://arxiv.org/abs/1709.07883}{{\tt
  arXiv:1709.07883}}].

\bibitem{Law:2020tsg}
Y.~T.~A. Law and M.~Zlotnikov, {\it {Massive Spinning Bosons on the Celestial
  Sphere}},  \href{http://arxiv.org/abs/2004.04309}{{\tt arXiv:2004.04309}}.

\bibitem{Johansson:2019dnu}
H.~Johansson and A.~Ochirov, {\it {Double copy for massive quantum particles
  with spin}},  {\em JHEP} {\bf 09} (2019) 040,
  [\href{http://arxiv.org/abs/1906.12292}{{\tt arXiv:1906.12292}}].

\bibitem{Porrati:2008ha}
M.~Porrati and R.~Rahman, {\it {A Model Independent Ultraviolet Cutoff for
  Theories with Charged Massive Higher Spin Fields}},  {\em Nucl. Phys. B} {\bf
  814} (2009) 370--404, [\href{http://arxiv.org/abs/0812.4254}{{\tt
  arXiv:0812.4254}}].

\bibitem{Rahman:2009zz}
R.~Rahman, {\it {Interacting massive higher spin fields}},  other thesis, 5,
  2009.

\bibitem{Caron-Huot:2016icg}
S.~Caron-Huot, Z.~Komargodski, A.~Sever, and A.~Zhiboedov, {\it {Strings from
  Massive Higher Spins: The Asymptotic Uniqueness of the Veneziano Amplitude}},
   {\em JHEP} {\bf 10} (2017) 026, [\href{http://arxiv.org/abs/1607.04253}{{\tt
  arXiv:1607.04253}}].

\bibitem{Afkhami-Jeddi:2018apj}
N.~Afkhami-Jeddi, S.~Kundu, and A.~Tajdini, {\it {A Bound on Massive Higher
  Spin Particles}},  {\em JHEP} {\bf 04} (2019) 056,
  [\href{http://arxiv.org/abs/1811.01952}{{\tt arXiv:1811.01952}}].

\end{thebibliography}\endgroup

\end{document}